\documentclass[floatfix,twocolumn,amssymb, amsmath,nobibnotes,aps,prb,superscriptaddress,showkeys]{revtex4-1}

\usepackage[T1]{fontenc}
\usepackage[utf8]{inputenc}
\usepackage{amssymb}
\usepackage{graphicx}
\usepackage{amsmath}
\usepackage{bm}
\usepackage{verbatim}
\usepackage{algorithm}
\usepackage{algorithmicx}
\usepackage{algpseudocode}
\usepackage{enumitem}
\usepackage{tabularx}
\usepackage{siunitx}
\usepackage[abs]{overpic}
\usepackage{pict2e}
\usepackage{ragged2e}
\usepackage[
bookmarksopen,
bookmarksdepth=2,
breaklinks=true
]{hyperref}
\usepackage{microtype}  
\usepackage{booktabs}
\usepackage{tabularx}
\newcolumntype{C}[1]{>{\centering\let\newline\\\arraybackslash\hspace{0pt}}m{#1}}
\newcolumntype{L}[1]{>{\hsize=#1\hsize\RaggedRight} X}

\PassOptionsToPackage{super}{natbib}
\newcommand{\ra}[1]{\renewcommand{\arraystretch}{#1}}
\bibpunct{}{}{,}{s}{}{;}
\setlist[itemize]{noitemsep}

\usepackage{xcolor}

\begin{document}

\title{Data-driven materials science: status, challenges and perspectives}

\author{L. Himanen}
\affiliation{Department of Applied Physics, Aalto University, P.O. Box 11100, 00076 Aalto, Espoo, Finland}

\author{A. Geurts}
\affiliation{Department of Applied Physics, Aalto University, P.O. Box 11100, 00076 Aalto, Espoo, Finland}
\affiliation{Department of Management Studies, Aalto University, P.O. Box 11100, 00076 Aalto, Espoo, Finland}
\affiliation{TNO, Netherlands Organization for Applied Scientific Research, Expertise Center for Strategy and Policy,  Anna van Beurenplein 1, 2595 DA The Hague}

\author{A. S. Foster}
\affiliation{Department of Applied Physics, Aalto University, P.O. Box 11100, 00076 Aalto, Espoo, Finland}
\affiliation{Graduate School Materials Science in Mainz, Staudinger Weg 9, 55128, Germany}
\affiliation{WPI Nano Life Science Institute (WPI-NanoLSI), Kanazawa University, Kakuma-machi, Kanazawa 920-1192, Japan}

\author{P. Rinke}
\email[Corresponding author: ]{patrick.rinke@aalto.fi}
\affiliation{Department of Applied Physics, Aalto University, P.O. Box 11100, 00076 Aalto, Espoo, Finland}
\affiliation{Theoretical Chemistry and Catalysis Research Centre, Technische Universit\"at M\"unchen, Lichtenbergstr.~4, D-85747 Garching, Germany}

\date{\today}

\begin{abstract}

Data-driven science is heralded as a new paradigm in materials science. In this field, data is the new resource, and knowledge is extracted from materials data sets that are too big or complex for traditional human reasoning - typically with the intent to discover new or improved materials or materials phenomena. Multiple factors, including the open science movement, national funding, and progress in information technology, have fueled its development. Such related tools as materials databases, machine learning, and high-throughput methods are now established as parts of the materials research toolset.
However, there are a variety of challenges that impede progress in data-driven materials science: data veracity, integration of experimental and computational data, data longevity, standardization, and the gap between industrial interests and academic efforts. 
In this perspective article, we discuss the historical development and current state of data-driven materials science, building from the early evolution of open science to the rapid expansion of materials data infrastructures. We also review key successes and challenges so far, providing a perspective on the future development of the field.
\end{abstract}



\maketitle

\section{Introduction}
In this perspective article, we review the current state of data-driven materials science with a focus on materials data infrastructures. \emph{Data-driven} invokes associations with big data, data management, open data and artificial intelligence (e.g. machine learning). The public debate of these terms is currently dominated by internet giants like Google, Amazon, and Facebook who also lead the technological development of data infrastructures, algorithms, and analysis tools. Compared to these e-commerce and social media developments, the field of data-driven \emph{materials science} is still under construction. By way of analogy, it is nonetheless still instructive to imagine a Materials ``Google'' -- the Materials Ultimate Search Engine (MUSE). In this article, we address what it takes to develop such a search tool for materials. 

\begin{figure}[h!]
  \centering
    \includegraphics[width=\columnwidth]{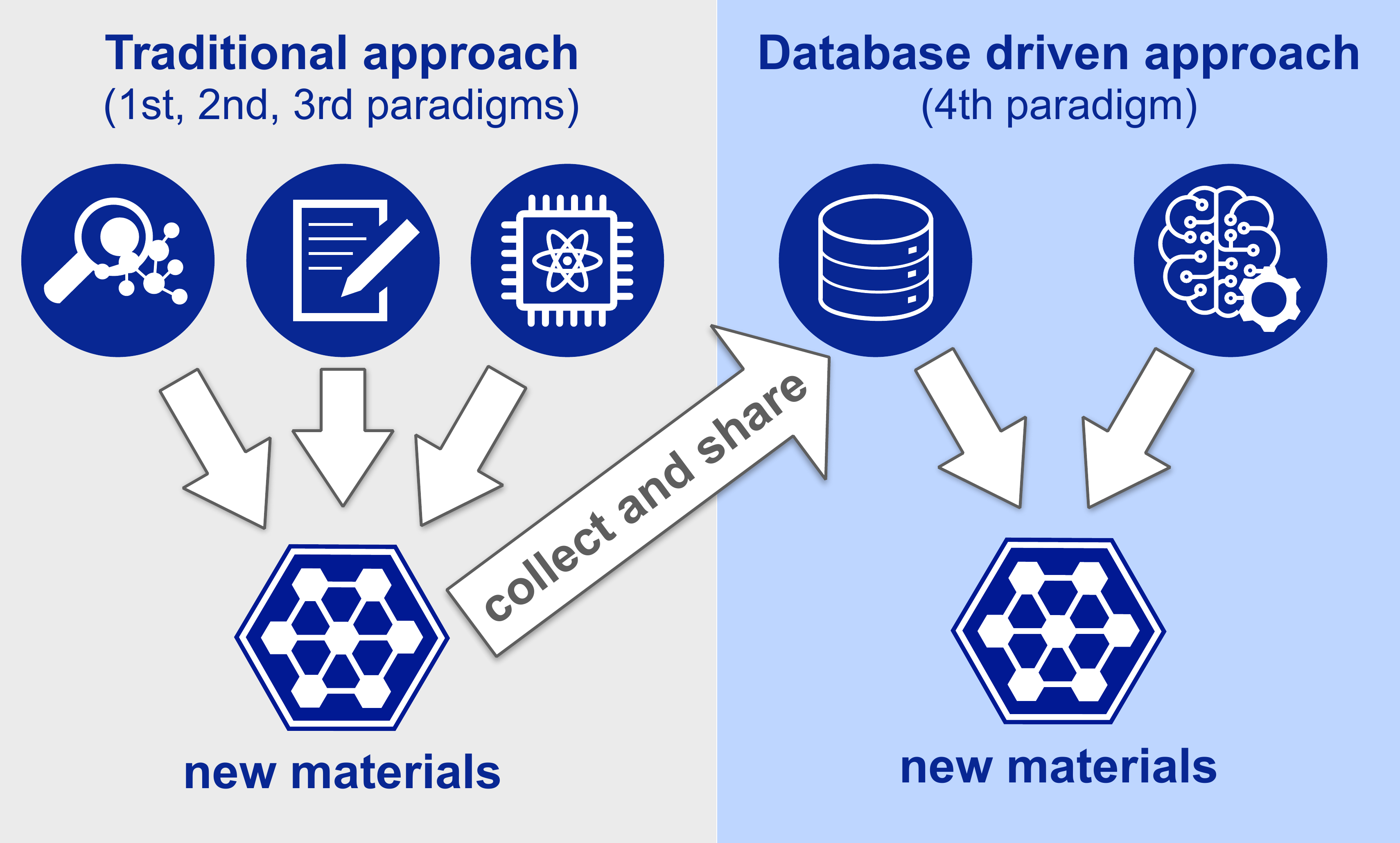}
  \caption{Materials discovery schematic. In the traditional approach, new materials are discovered by experimentation, theory, or computation (also referred to as 1st, 2nd, and 3rd paradigms and symbolized by the three icons at the top of the left panel). In the 4th paradigm of data-driven materials science, available data is gathered in data infrastructures, and machine learning approaches discover new materials.}
\label{fig:datadrivenmaterialsscience}
\end{figure}

Materials science, the study of the characteristics and applications of materials, is a well established discipline that combines chemistry, physics, and engineering research. Materials scientists frequently dream of designing new materials from scratch for use in society\cite{Ceder}. However,  instead of finding new materials using the MUSE, they discover new materials through conventional experimental, theoretical, or computational research (see left panel of Fig.~\ref{fig:datadrivenmaterialsscience}). This pipeline through which new materials are discovered, designed, developed, manufactured, and deployed remains slow, costly, and highly inefficient: \emph{By the time a new material comes to market, the patent protection of the original invention is at the end of its tenure, and proprietary advantage is lost}\cite{Eagar:1995} (see also Ref.~\onlinecite{McKinsey}). By applying data science to materials research, we now have a way to accelerate the materials value chain from discovery to deployment. 

Data science has developed out of the growing demand for open science combined with 
the meteoric rise of AI and machine learning. As these innovative technologies allow ever-larger datasets to be processed and hidden correlations to be unveiled, data-driven science is  emerging as the \emph{fourth scientific paradigm} \cite{fourthparadigm,Agrawala} (cf. Fig.~\ref{fig:datadrivenmaterialsscience}) following the first three eras of experimentally, theoretically, and computationally propelled scientific discoveries. Often connected to the \emph{fourth industrial revolution}\cite{fourthindustrialrevolution} or the \emph{second machine age},\cite{machineage} such data-driven approaches permeate science, business, politics, and even social life. Since materials innovation is a critical, well-recognized driver of economic development and societal progress, it is important that new trends, such as data science, are embraced if they have the potential to advance the field.

\textit{Data-driven materials science} and \textit{materials informatics} are umbrella terms for the scientific practice of systematically extracting knowledge from materials datasets. This practice differs from traditional scientific approaches in materials research by the volume of processed data and the more automated way information is extracted (cf. Fig.~\ref{fig:datadrivenmaterialsscience}),  for example,  through  the  use  of  machine learning (see Ref.~\onlinecite{RAJAN200538, Rajan:2015kf, Agrawala, Liu2017, Zdeborova2017, nutshell, Ramprasad2017,Ward2017,Rampi_review,Zunger:2018,Butler2018,Gubernatis:2018gi,Goldsmith:2018bb, Takahashi:2016hr,tms,Bartok/etal:2017, C9TA02356A} for recent review articles on machine learning in materials science). In our MUSE analogy, this would be the \emph{search} and \emph{find} part. In addition to data processing and data analysis tools, data-driven materials science also requires physical infrastructures that host and preserve that data. These would be the \emph{data storage} part of our MUSE example, which, as physical infrastructures, require dedicated community efforts and sustained investment to become and remain operational.

Stakeholders in academia, industry, governments, and the public attach different meanings and expectations to data-driven materials science. The actual material \emph{science} is carried out in academia and research and development (R\&D) departments in industry. Scientists at universities and companies not only produce materials data that could then be stored in data facilities, they are also the primary user group of materials data infrastructures. In the wake of digitalization, industry has a further interest in digitizing materials data and incorporating data-driven materials science into their value chain. Policy makers and governmental or private funding agencies may have an interest in promoting open science data and can stir scientific developments through policy and funding decisions. The general public benefits from materials science by quality-of-life enhancement through new products and technologies. They have an indirect interest in data-driven materials science as a means to accelerate innovations and follow developments in science and open data in the media. Together these stakeholders form an ecosystem of mutual benefit. The vitality of this ecosystem is crucial for the success and the longevity of data-driven materials science.

In this article, we embed our perspective in the emerging field of data-driven materials science in the context of the open science movement, which has shaped the philosophy and design of several materials science data infrastructures. We discuss how these infrastructures grew historically from simple databases into data centres that then progressed into materials discovery platforms, and we detail the current state of data infrastructure. A list of current challenges provides the gateway to the second part of this article, in which we delve deeper into data organization, acquisition, quality, and machine learning. We conclude with an industrial perspective that  addresses the future and longevity of materials data infrastructures.

\section{Open Science movement}\label{open_science}
Many of the fundamental aspects of data-driven materials science are built upon the key elements of the Open Science movement. The European Commission outlines \cite{Commission:kGSfz0hY} Open Science as \emph{ ``...a new approach to the scientific process based on cooperative work and new ways of diffusing knowledge by using digital technologies and new collaborative tools. ''} Here we reflect on those aspects of Open Science that are particularly relevant to the birth and future of data-driven materials science. 

Openness in science was initially curtailed by the prestige wars between the patrons of early scientists and their associated, convoluted encryption schemes \cite{Regazzi:1985241}. Once more professional scientific practice developed, scientists embraced the idea of accessibility of research as a cornerstone of progress, and this has been generally mandated by public policy. As early as 1710 in the UK, the Copyright Act endowed the ownership of copyright to authors rather than publishers, encouraging authors to deposit manuscripts into national libraries to make them publicly accessible. In addition to accessibility, public accountability and scientific reproducibility have remained powerful driving forces in the way science has been conducted and disseminated, and significant deviations from these norms are of great concern to the community \cite{Fanelli:2018je}. More recently in the 1990s, the development of the internet transformed this debate as it became possible to make nearly all aspects of the scientific research process easily accessible, from preliminary data to final publications. While arguments over fair allocation of rewards for scientific achievement versus full and early research dissemination remain challenging \cite{Tennant:2016hq}, and intellectual property management regularly introduces conflicts \cite{Esanu:2003cw}, the era of Open Science \cite{open_science,foster} (or indeed Open Innovation \cite{chesbrough}) is here to stay and contributes to scientific advancement overall \cite{McKiernan:2016iz}. Open-access journals and data, and open-source software have significant impacts on the Open Science movement.

\subsection{The rise of Open Access publishing}\label{open_pub}
Building on the foundations of the very first online journals, websites like arXiv (established in 1991) took the first steps in providing Open Access to scientific publications. As more content became available online, and the need for physical copies of journals in libraries rapidly diminished, many expected a significant reduction in the cost of journal subscriptions. When this did not happen, it catalyzed the Open Access movement and other alternatives to conventional scientific publishing practices. At present, over 50\% of newly published articles are Open Access, and conservative estimates place achievement of complete Open Access by 2040 \cite{oa_when,Piwowar:2018bj}. Current Open Access approaches tend to fall into two classes (or hybrids thereof \cite{Piwowar:2018bj}): gold, where the article is freely available at the point of publication; and green, where the authors can deposit the article in a public repository, for example, at their home institution. Some publishers require an embargo period before deposition in a public repository, but there is little evidence in terms of publisher income to support the existence of such embargoes \cite{Tennant:2016hq}. Many funding agencies have embraced Open Access publishing as a way to improve public transparency and accountability, and these agencies have made it a condition for support - this includes all European Union funding for 2020 and beyond \cite{Schiltz:2018bp}. As such, Open Access is at the heart of the Open Science movement and certainly overlaps with one of the critical developments in data-driven science, Open Data.  

\subsection{Open Access data}\label{open_data}
The initiative to make data Open Access can be traced to efforts to establish scientific global data centres in the 1950s \cite{wdc}, largely as a way to store data long term and make it internationally accessible - all data was fully available for the cost of printing and delivery. Following this change, demands for scientific data sharing continued to rise, especially after the development of the internet and the tantalizing prospect of easy upload and download of data globally. 

While many scientists were quick to embrace this, it took a decade for Open Data to appear as a clear objective and topic for scientific policy. In 2004, science ministers of most developed countries signed an agreement that all publicly funded archive data should be made available, with the guidelines for this following in 2007 \cite{oecd_open2007}. As is often the case, the scientific communities themselves were ahead of policy changes, and many bespoke scientific databases had already proliferated, providing data repositories in almost every field across the globe. There are now thousands of them, and finding useful ways to search for a relevant repository, let alone data within it, requires serious effort \cite{re3data}. 

Motivation to make this effort is increasing rapidly, with many journals and funding agencies demanding the availability of data tied to publication or grants. Contributing to many aspects of the Open Science initiative, the Public Library of Science \cite{plos} has pioneered this development, with a clear policy on data sharing for its publications and likely rejection if policies are not followed. Other major publishers have also been active, with at least the creation of specific Open Data journals \cite{sdata}, policies \cite{elsevdata}, and collaboration with Open Data initiatives \cite{codata}. Many funding agencies now insist on a data management plan with all submissions, and this plan must give a detailed account of how data will be stored, secured, and shared - with particular attention to the Open Science rules of the agency in question. 

In an attempt to provide unifying guidelines for the widely varying groups interested in Open Data and to aid in data management development, the FAIR Data Principles were established \cite{Wilkinson:2016dn,NomadFair}. These principles have been adopted by several major players in global data management (see Table \ref{tbl:databases}). The ideas behind making data searchable, accessible, flexible, and reusable at the core of FAIR are also the concepts that make the power of data-driven science actually attainable.

\subsection{Open-source software for science}\label{open_soft}
The development of open-source software entails the final element of the Open Science movement. Its development started in parallel with the earliest computing hardware efforts, with nearly all software freely available in the public domain as part of large academic and corporate collaborations. Since the relative cost of software compared to hardware has increased, this openness began to steadily decline until the early 1980s with the launch of the GNU project and the parallel explosion of Linux and the internet in the early 1990s. This provided a powerful platform and toolset for the collaborative development of software that could then be freely downloaded, culminating in the active open-source movement in 1997 \cite{KELTY:kl}. In particular, it suited the kind of focused, rapidly changing software that characterizes nearly all scientific applications. 

In 2005, the creation (by Linus Torvalds) and rapid adoption of Git as a distributed revision control system, closely followed by hosting site GitHub, put the seal on the standard approach for open-source scientific software development that remains to this day. It became possible to manage updates to codes from a large development team, while providing a platform for feedback, bug notification, and feature requests from users. It is now possible to find Open Source software for nearly every aspect of a scientific project \cite{opensoft}, from electronic lab notebooks \cite{Kwok:2018cw}, experimental toolsets \cite{Horcas:2007jp}, and simulation packages \cite{Plimpton:1995fc}, to machine learning libraries \cite{scikit-learn} and online collaborative writing sites \cite{overleaf}. With freely accessible data, Open Access publications explaining the science behind it, and a wealth of open-source software to mine it, the way is clear for innovative data-driven science.

\section{Materials data infrastructures \label{infra}}

Having established the context for Open Science, we next review the emergence of materials data infrastructures that collect, host, and provide materials data to stakeholders. We first reflect on early digital materials infrastructures before discussing the current state.

\begin{figure*}[t]
  \centering
    \includegraphics[width=\textwidth]{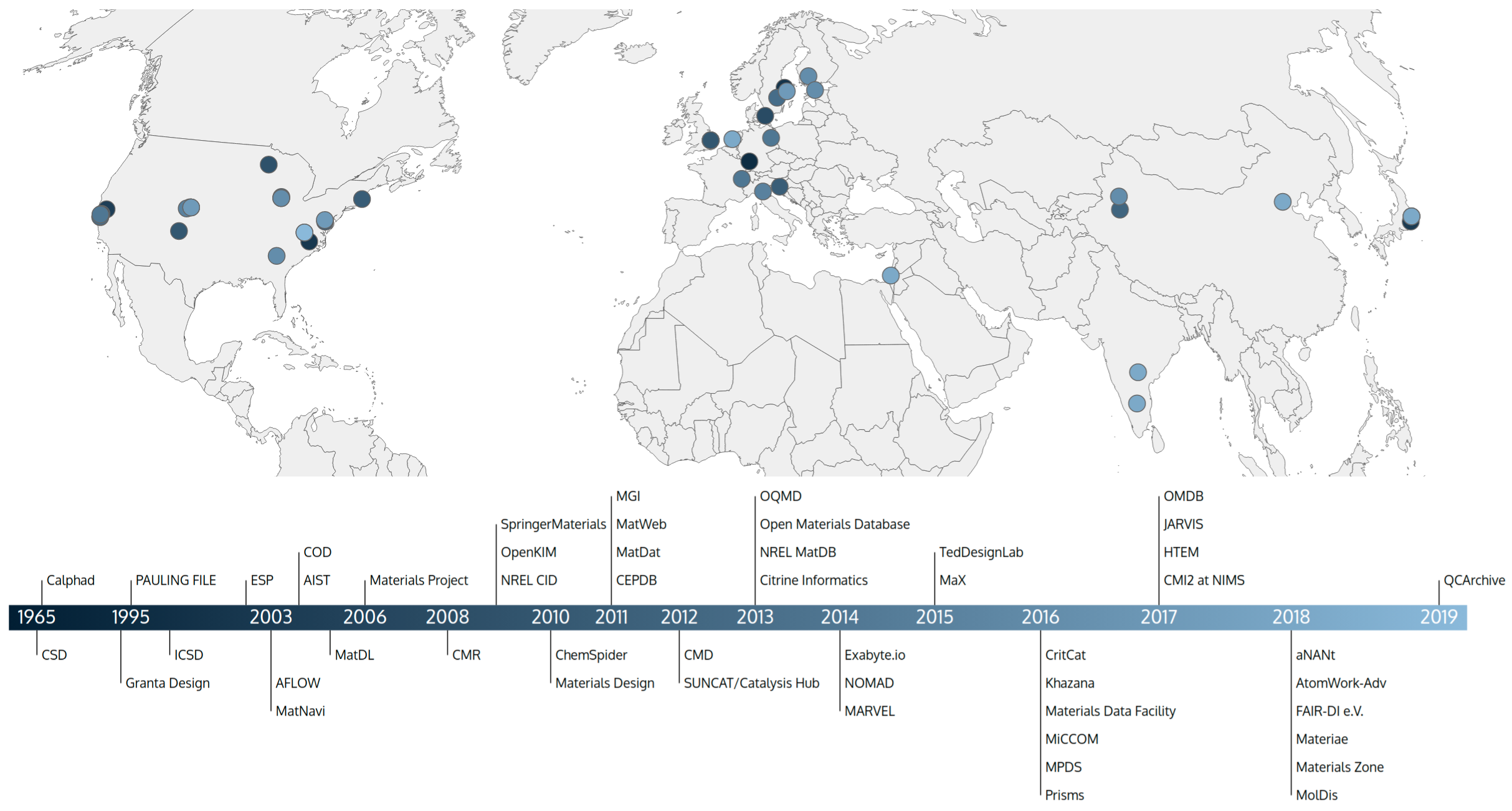}
  \caption{Timeline and geographic distribution of materials data infrastructures and companies. The colour of the dots represents the time of establishment. The map shows that historically more centers have emerged in the U.S. and Europe, with Asia catching up over time. In addition, the U.S. has a higher renewal rate than Europe, as can be seen in the larger number of ligher colored dots. CSD: Cambridge Structural Database, ICSD: Inorganic Crystal Structure Database, ESP: Electronic Structure Project, AFLOW: Automatic-Flow for Materials Discovery, AIST: National Institute of Advanced Industrial Science and Technology Databases, COD: Crystallography Open Database, MatDL: Materials Digital Library, CMR: Computational Materials Repository, NREL CID: NREL Center for Inverse Design, CEPDB: The Clean Energy Project Database, MGI: Materials Genome Initiative, CMD: Computational Materials Network, OQMD: Open Quantum Materials Database, NOMAD: Novel Materials Discovery Laboratory, MaX: Materials Design at the Exascale, MICCOM: Midwest Integrated Center for Computational Materials, MPDS: Materials Platform for Data Science, CMI2: Center for Materials Research by Information Integration, HTEM: High Throughput Experimental Materials Database, JARVIS: Joint Automated Repository for Various Integrated Simulations, OMDB: Organic Materials Database, QCArchive: The Quantum Chemistry Archive}
\label{fig:map_of_centres}
\end{figure*}

\subsection{Development of materials infrastructures}
\label{sec:history_of_time}

The increasing capabilities of first-principles methods - and the increasing capabilities of computational science in general - have accelerated materials researchers interest for new, computer-based pathways to materials discovery and design - better, faster, and cheaper than ever before. Perhaps one of the first attempts to use materials information in a different and more efficient way was the development of the Calculation of Phase Diagrams (CALPHAD, 1970s) method and database, in which multiple calculations of phase diagrams were put in a centralized database to speed up the design and development of new alloys\cite{Calphad}. In the 1990s, the increasing capability to collect, store and analyse ``big data'' led researchers to explore the potential of data-science in  scientific research (for more information, see\cite{fourthparadigm}). With these innovative ideas up in the air, material scientists at the Massachusetts Institute of Technology (MIT) developed tools to predict the properties of materials from datasets \cite{ceder1997}. Around the same time, researchers at the Technical University of Denmark demonstrated the potential of evolutionary algorithms in finding materials with specific properties \cite{PRL2002}, or to use high-throughput screening for candidate materials with key parameters to narrow down the number of required experiments \cite{Norskov,Catapp,Greeley}. The researchers at MIT even envisioned how with such computational tools a ``virtual materials laboratory'' could be build, in which new materials are designed and tested based on computer calculations\cite{ceder1997}. These ideas eventually led to the launch of a curated database that is now called the Materials Project\cite{Ceder2006,materialsproject}. This Open Access (see Sec.\ref{open_data}) database would use high-throughput computing to uncover the properties of all known inorganic materials and enable future researchers to find appropriate materials through interactive exploration and data mining \cite{materialsproject,Jain2011}. 

\begin{figure}[h!]
  {\centering
    \includegraphics[width=\columnwidth]{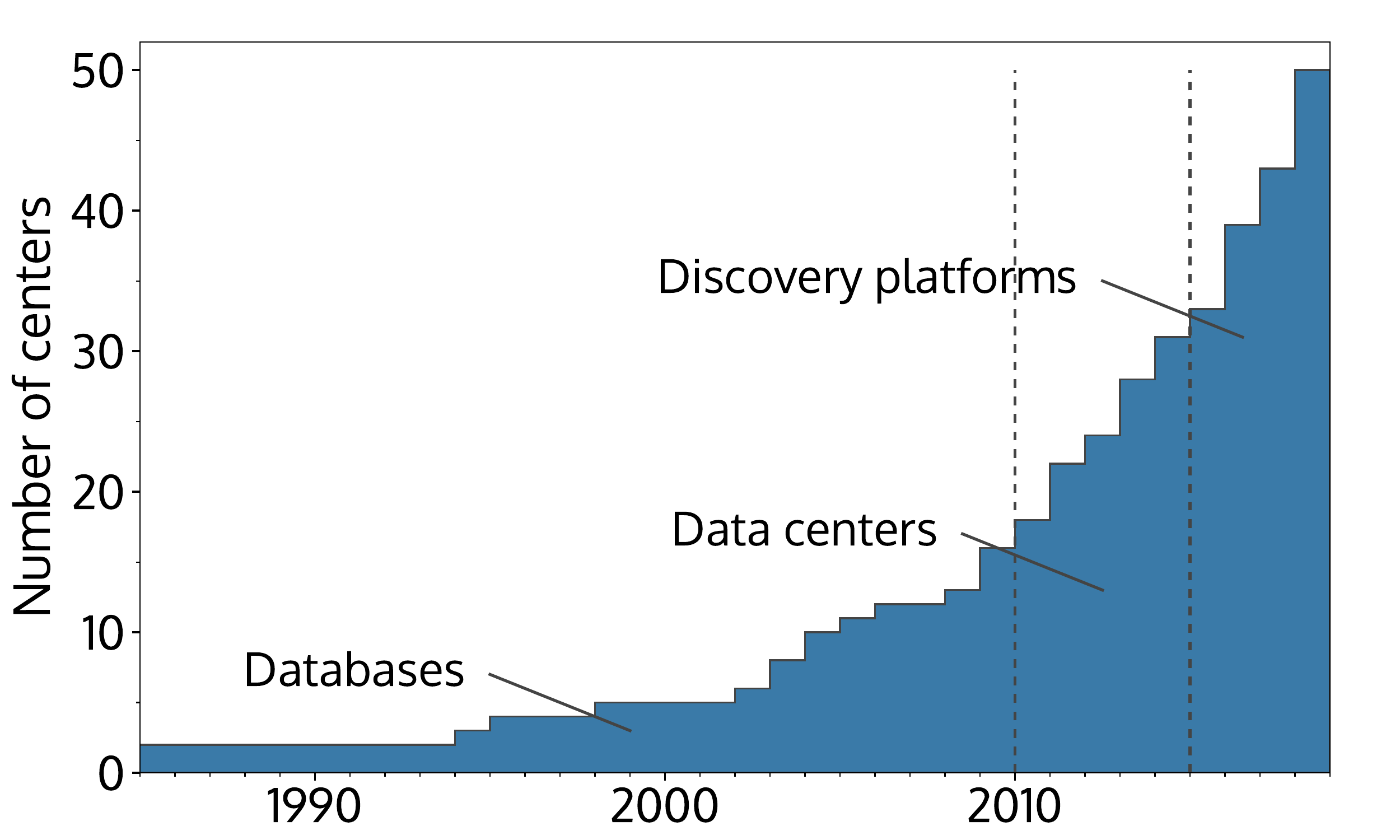}}
  \caption{Number of materials informatics projects and infrastructures as function of time (see Fig.~\ref{fig:map_of_centres} and Tab.~\ref{tbl:databases} for details on individual projects and infrastructures). We divide the time axis into three periods that reflect the evolution of the data infrastructures (see text for details).}
\label{fig:centre_time_evolution}
\end{figure}

As big data and data science became increasingly fashionable, the US government announced the launch of the Materials Genome Initiative (MGI) in 2011\cite{materialsgenome}. This initiative emphasized the usefulness of data informatics for materials discovery and design. As similar efforts were launched around the world promoting the availability and accessibility of digital data in science, a trend was set and a new paradigm of materials science emerged\cite{Agrawala}: data-driven materials science. Set to reduce time and investment needed to support the typical 10-20 year research-development-commercialization cycle for new materials, more and more Open Access materials data initiatives opened worldwide, as illustrated by Figs.~\ref{fig:map_of_centres} and \ref{fig:centre_time_evolution}. 

Most of the early materials data initiatives started  as \emph{databases} that hosted data and offered search functionality with the idea to encourage materials scientists to share their data with a larger community. The launch of the Materials Genome Initiative became a defining moment in data-driven materials science (see Fig.~\ref{fig:centre_time_evolution}) as databases evolved into \emph{data centres} that offered rudimentary materials and data analysis services. The emerging interest around data mining and AI made materials scientists increasingly eager to use such algorithms in their research. As a result, the focus of most centres transitioned to developing workflows that would enable scientists to search, mine, and query the databases. This marks another turning point in the history of data-driven materials science, with infrastructures becoming \emph{materials discovery platforms} (see Fig.~\ref{fig:centre_time_evolution}), whose self-declared mission is to facilitate the discovery of novel materials.

The distinction between \emph{databases}, \emph{data centers} and \emph{materials discovery platforms} introduced in the previous paragraph is based on the loose definitions given in the paragraph. The terminology reflects our impression of the evolution of materials data infrastructures and provides a simple classification scheme to distinguish different infrastructure types. For the remainder of the article, we will use  \emph{materials data infrastructure} as the most general and encompassing term to refer to either of the three types.

The spillover effect from data science to materials science is currently boosting the emerging field of data driven materials science or materials informatics\cite{Agrawala}. The computational possibilities of machines to analyse and detect patterns in data has created a new feedback loop in the relationship between hypothesis and experiment, which facilitates the next step to mix human trial-and-error experimental and computational research with ``artificial intuition'' (or: to use data mining tools to approach human-like intuition to suggest candidate materials that are further refined via computational and experimental research). 

Big data and data science are also prevalent in other scientific fields. In chemistry, databases emerged earlier than in materials science \cite{Hann/Green:1999,Noordik:2004,Willet:2011}, as exemplified by Chemical Abstracts Service (CAS), the principal chemical database provider \cite{Baker/etal:1980,Weisgerber:1997} whose first database was created in 1965 \cite{Weisgerber:1997,Leiter/etal:1965}. Carefully produced and curated data sets were essential for developments in quantum chemistry \cite{Curtiss/etal:1991,Curtiss/etal:1997,Goerigk/Grimme:2011,Mata/Suhm:2017}. In particle physics, the CERN Open Data Portal\cite{CERN_opendata} offers more than 1 petabyte of open data for research conducted at their facilities. In biology, a variety of databases and metadatabases store biological information, for example ConsensusPathDB\cite{ConsensusPathDB} for human protein-protein, genetic, metabolic, signaling, gene regulatory, and drug-target interactions; the protein data bank \cite{ProteinDataBank} that houses three-dimensional structural data of large biological molecules; and the International Nucleotide Sequence Database Collaboration\cite{INSDC,Brunak1333} that collects and disseminates DNA and RNA sequences. In this perspective article, our focus is on materials science, but it is clear that the ``4th scientific paradigm''\cite{fourthparadigm} is emerging in other fields as well. 

\subsection{Overview of current materials infrastructures}


Figure \ref{fig:centre_time_evolution} depicts a clear rise of active materials infrastructures, many of which have developed into very mature and stable services used in everyday research processes \cite{aflow_application, mp_application, cmr_application, marcaki}. Table \ref{tbl:databases} shows a summary of the most prominent materials discovery platforms in existence today. As these platforms have matured, the range of different services they provide has grown (for another perspective on the components of materials data infrastructures see \cite{tms}), and Table \ref{tbl:features} shows their features. Data infrastructures that have emerged at the time of submission of this article, such as, e.g., QCArchive \cite{QCarchive} have not yet been included in Tabs.~\ref{tbl:databases} and \ref{tbl:features}.


Perhaps the most important service that a data platform has to offer is an efficient distribution channel for the data stored within. Often the data is accessible through a webpage that its clients can access online. This has the lowest adoption barrier since no additional software is needed and the data can be explored visually through a browser. Examples of such services include the NOMAD Encyclopedia\cite{encurl}, AFLOWlib\cite{aflowliburl}, the Materials Project\cite{mpurl}, and the Materials Cloud\cite{mcurl}. A browser-based method is, however, rarely useful for materials informatics applications, which require automated access to large volumes of data. To facilitate access to large data volumes, it is typical to offer an application programming interface (API) to users to enable automatic data crawling. This is often done by defining a Representational State Transfer (REST) or GraphQL interface to the data\cite{nomadrest, aflowrest, mprest, cathuburl}. These interfaces allow automated access through programmable queries. Another way, as adopted by OQMD\cite{oqmd} for example, is to offer an offline version of the database as a direct download to users. Offline access provides the most flexibility and performance but typically requires knowledge on how to interact with the underlying database with Structured Query Language (SQL) or object-relational-mapping (ORM). That said, a full download is not practical for large data volumes.

\begin{table*}[p!]
\scriptsize
\centering
\ra{1.6}
\begin{tabularx}{1\textwidth}{L{0.25} L{0.25} L{0.3} L{1.0} l}
\toprule
Name & Website & Contact & Overview & Ref. \\
\midrule
AFLOW & aflowlib.org & Stefano Curtarolo, Duke University & Computational data consisting of 2,118,033 material compounds and 281,698,389 calculated properties with focus on inorganic crystal structures. Incorporates multiple computational modules for automating high-throughput first principles calculations. & \citenum{aflowlib, aflowliburl} \\
Computational Materials Repository & cmr.fysik.dtu.dk & Kristian Thygesen and Karsten Jacobsen, DTU & Computational datasets from a diverse set of applications. Data creation and analysis with the Atomic Simulation Environment (ASE). & \citenum{cmr2, cmr, cmrarticle} \\
Crystallography Open Database & crystallo\-graphy.net & 
& Open-access collection of crystal structures of organic, inorganic, metal-organics compounds and minerals, excluding biopolymers. & \citenum{cod,codurl} \\
HTEM & htem.nrel.gov & Caleb Phillips and Andriy Zakutayev, NREL & Properties of thin films synthesized using combinatorial methods. Contains 57597 thin film samples, across a wide range of materials (oxides, nitrides, sulfide, intermetallics). & \citenum{htem,htemurl} \\
Khazana & khazana\-.gatech.edu & Rampi Ramprasad, Georgia Institute of Technology & Platform to store structure and property data created by atomistic simulations, and tools to design materials by learning from the data. Tools include Polymer Genome and AGNI. & \citenum{khazana, polymergenome, agni} \\
MARVEL NCCR & nccr-marvel.ch & Nicola Marzari, EPFL & Materials informatics platform for data-driven high-throughput quantum simulations. Data available at materialscloud.org, powered by the AiiDA-infrastructure. & \citenum{mcurl} \\
Materials Data Facility (MDF) & materialsdata\-facility.org & Ben Blaiszik and Ian Foster, University of Chicago & Data publication network for computational and experimental datasets. Data exploration through the Forge python package. & \citenum{mdf,mdfurl} \\
Materials Project & materials\-project.org & Kristin Persson, LBNL & Online platform for materials exploration containing data of 86,680 inorganic compounds, 21,954 molecules and 530,243 nanoporous materials. Develops various open-source software libraries, including pymatgen, custodian, FireWorks, and atomate. & \citenum{Jain:2013ku,mpurl} \\
MatNavi/NIMS & mits.nims.go.jp  & Yibin Xu, NIMS & An integrated material database system comprising ten databases, four application systems and the NIMS Structural Datasheet Online. & \citenum{MatNavi} \\
NOMAD CoE & nomad-coe.eu & Matthias Scheffler, FHI/Max Planck Society & Provides storage for full input and output files of all important computational materials science codes, with multiple big-data services built on top. Contains over 50,236,539 total energy calculations. & \citenum{nomad, repourl} \\
Organic Materials Database & omdb.mathub.io & Alexander Balatsky, Nordita & Open access electronic structure database for 3-dimensional organic crystals. Contains approximately 24 000 materials. & \citenum{omdb1, omdb2} \\
Open Quantum Materials Database & oqmd.org & Chris Wolverton, Northwestern University & Database of DFT-calculated thermodynamic and structural properties with focus on inorganic crystal structures. Contains 563,247 entries with support for full download and advanced usage through the qmpy python package. & \citenum{oqmd,oqmdurl} \\
Open Materials Database & open\-materialsdb.se & Rickard Armiento, Linköping University & Computational database primarily based on structures from the Crystallography Open Database. Data creation and analysis with High-Throughput Toolkit (httk). & \citenum{omdurl, httkurl} \\
SUNCAT & suncat.stanford\-.edu & Thomas Francisco Jaramillo, SLAC/Stanford University & Materials informatics center for atomic-scale design of catalysts. Online tools and computational results for 112,157 surface reactions and barriers available at catalysis-hub.org. & \citenum{suncaturl,cathuburl} \\
\midrule
Citrine Informatics & citrine.io & Bryce Meredig and Greg Mulholland & A materials informatics platform combining data infrastructure and AI. Open database and analytics platform for material and chemical information available at the Citrination platform: citrination.com. & \citenum{citrine,citrineurl} \\
Exabyte.io & exabyte.io & Timur Bazhirov & Cloud-based modelling platform for materials informatics. & \citenum{exabyteurl,exabyte} \\
Granta Design & granta\-design.com & Mike Ashby and David Cebon & R\&D organization offering data, tools and expertise for materials design. & \citenum{grantaurl} \\
Materials Design & materials\-design.com & Clive M. Freeman, Erich Wimmer and Stephen J. Mumby & Software products and services for chemical, metallurgical, electronic, polymeric, and materials science research applications. & \citenum{md1,md2,mdurl} \\
Materials Platform for Data Science & mpds.io & Evgeny Blokhin & Online edition of the PAULING FILE with focus on curated experimental data for inorganic materials. & \citenum{mpds1, mpds2} \\
MaterialsZone & materials.zone & Assaf Anderson and Barak Sela & Provides a notebook-based materials informatics environment together with experimental data. & \citenum{mzurl} \\
SpringerMaterials & materials.\-springer.com & Michael Klinge & Curated data covering multiple material classes, property types, and applications. A set of advanced functionalities for visualizing and analyzing data provided through SpringerMaterials Interactive. & \citenum{springermaterials} \\
\bottomrule
\end{tabularx}
\caption{List of current major materials data infrastructures. The entries are divided into non-commercial (top) and commercial (bottom). Note that some platforms are named after the leading research project and may host multiple services under different names. As contact person we listed the director(s) of each infrastructure, in such cases, where they were clearly identifiable. Data volume numbers reflect the state in April 2019.}
\label{tbl:databases}
\end{table*} 
\begin{table*}[t]
\footnotesize
\centering
\ra{1.6}
\begin{tabularx}{1\textwidth}{XC{15mm}C{15mm}C{15mm}C{15mm}C{15mm}C{15mm}C{15mm}}
\toprule
 & Open Access & Comp. data & Exp. data & Data upload (DOIs) & Workflow management tools & Web API & Data analysis tools \\
\midrule
AFLOW & \checkmark & \checkmark &  &  & \checkmark & \checkmark & \checkmark \\
Computational Materials Repository & \checkmark & \checkmark &  &  & \checkmark &  & \checkmark \\
Crystallography Open Database & \checkmark & \checkmark & \checkmark & \checkmark &  &  &  \\
HTEM & \checkmark &  & \checkmark  & \checkmark &  & \checkmark & \checkmark   \\
Khazana & \checkmark & \checkmark & \checkmark  &  &  &  & \checkmark   \\
MARVEL NCCR & \checkmark & \checkmark &  & \checkmark & \checkmark &  & \checkmark \\
Materials Data Facility (MDF) & \checkmark & \checkmark & \checkmark & \checkmark (DOI)* &  & \checkmark &  \\
Materials Project & \checkmark & \checkmark &  &  & \checkmark & \checkmark & \checkmark \\
MatNavi/NIMS & \checkmark & \checkmark & \checkmark &  &  &  & \checkmark \\
NOMAD CoE & \checkmark & \checkmark &  & \checkmark (DOI) &  & \checkmark & \checkmark \\
Organic Materials Database & \checkmark & \checkmark &  &  &  &  & \checkmark \\
Open Quantum Materials Database & \checkmark & \checkmark &  &  &  &  & \checkmark \\
Open Materials Database & \checkmark & \checkmark &  & \checkmark & \checkmark & \checkmark & \checkmark \\
SUNCAT & \checkmark & \checkmark &  &  &  & \checkmark & \checkmark \\
\midrule
Citrine Informatics & \checkmark** & \checkmark & \checkmark & \checkmark &  & \checkmark & \checkmark \\
Exabyte.io &  &  &  &  &  & \checkmark & \checkmark \\
Granta Design &  & \checkmark & \checkmark &  &  &  & \checkmark \\
Materials Design & & \checkmark & \checkmark &  &  &  & \checkmark \\
Materials Platform for Data Science & \checkmark*** & \checkmark & \checkmark &  &  & \checkmark & \checkmark \\
MaterialsZone &  &  & \checkmark &  &  & & \checkmark \\
SpringerMaterials & & & \checkmark &  &  &  & \checkmark \\
\bottomrule
\end{tabularx}
\caption{Services provided by the selected materials data infrastructures. Open Access: provides partial or full free access to data. Computational data: contains data originating from software simulations. Experimental data: contains data originating from experiments. Data upload: Allows upload of own data, with the possibility of issuing Digital Object Identifiers (DOIs). Workflow management tools: provides or collaborates in the development of open-source software tools for workflow management. Web API: data can be accessed remotely with automated scripts. Data analysis tools: provides online or offline data analysis tools, including machine learning. \\ \footnotesize{*Upload requires access to private/institutional storage space.} \\ \footnotesize{**Open Access to a subset of data.} \\
\footnotesize{***Open Access to limited set of materials properties.}}
\label{tbl:features}
\end{table*}

As the amounts of data produced by materials science increases, a practical concern over long-term storage of this data is emerging. There is also increasing pressure from funding agencies and other institutions to ensure the correct and safe long-term storage of data. To answer this demand, some data infrastructures now provide data storage services for materials data. Currently Springer-Nature lists two recommended data repositories for materials science\cite{naturerecommended}: the NOMAD Repository\cite{repourl} and the Materials Cloud\cite{matcloud}. Both of these free services are for computational materials data, accept uploads from any source, and guarantee data storage for at least 10 years after data deposition. Often the data volumes in experimental studies, especially in imaging, far outnumber computational efforts. For instance, electron microscopes can easily generate tens of gigabytes of data in a day of operation\cite{Kalidindi:2015kn}. Because of this higher volume, it is much more challenging to organize central and free data storage for experimental data. Instead the storage space is provided by the host university or laboratory, as in the case of the Materials Data Facility\cite{mdf}, which is a collaboration between US universities and research centres. In addition to the materials-science-specific storage solution, there are also free solutions to store generic scientific data, such as Zenodo\cite{zenodo}, Dryad\cite{dryad}, Figshare\cite{figshare}, and Dataverse\cite{dataverse}.

The online analytics tools\cite{nomadanalytics, mcurl, exabyte} provided by data infrastructures are fairly modern additions that have emerged from the rise and popularity of interactive browser content and notebook-based environments, such as the Jupyter notebook\cite{jupyter}. These online tools range from simple tutorials to realistic materials property prediction and materials discovery through machine learning. They can be used without local hardware or software resources and have therefore become an important channel for dissemination and learning. Some platforms also participate in the development of Open Source software (see Sec. \ref{open_soft}) libraries for performing offline data analysis on materials data\cite{ase, pymatgen, qmpy, nomadlab}. Such libraries have high reuse value for scientists working with materials data and, through  Open Source distribution and contribution mechanisms, can remain in active use beyond the lifetime of individual projects.

The value of materials data has also been recognized by materials informatics companies. We have included a selection of these companies in Tables \ref{tbl:databases} and \ref{tbl:features}. A major selling point of these companies is the access they provide to privately owned, highly curated materials property data that is inherently valuable in R\&D. In sufficiently large quantities, this kind of materials data can help firms immensely in selecting optimal materials for products, without having to spend additional expenses on building their own research infrastructure. Another recently emerging business model revolves around selling access to software environments with a Software as a Service (SaaS) model. In this model, companies offer on-demand access to preconfigured cloud-based environments for materials informatics. Such services can be valuable for companies and research laboratories because they can be used according to current demand, do not require large one-off investments in hardware, and do not require specialized skills in software configuration and system management.

\subsection{Current Challenges}
\label{sec:challenges}

\begin{figure}[h!]
  \centering
    \includegraphics[width=\columnwidth, ]{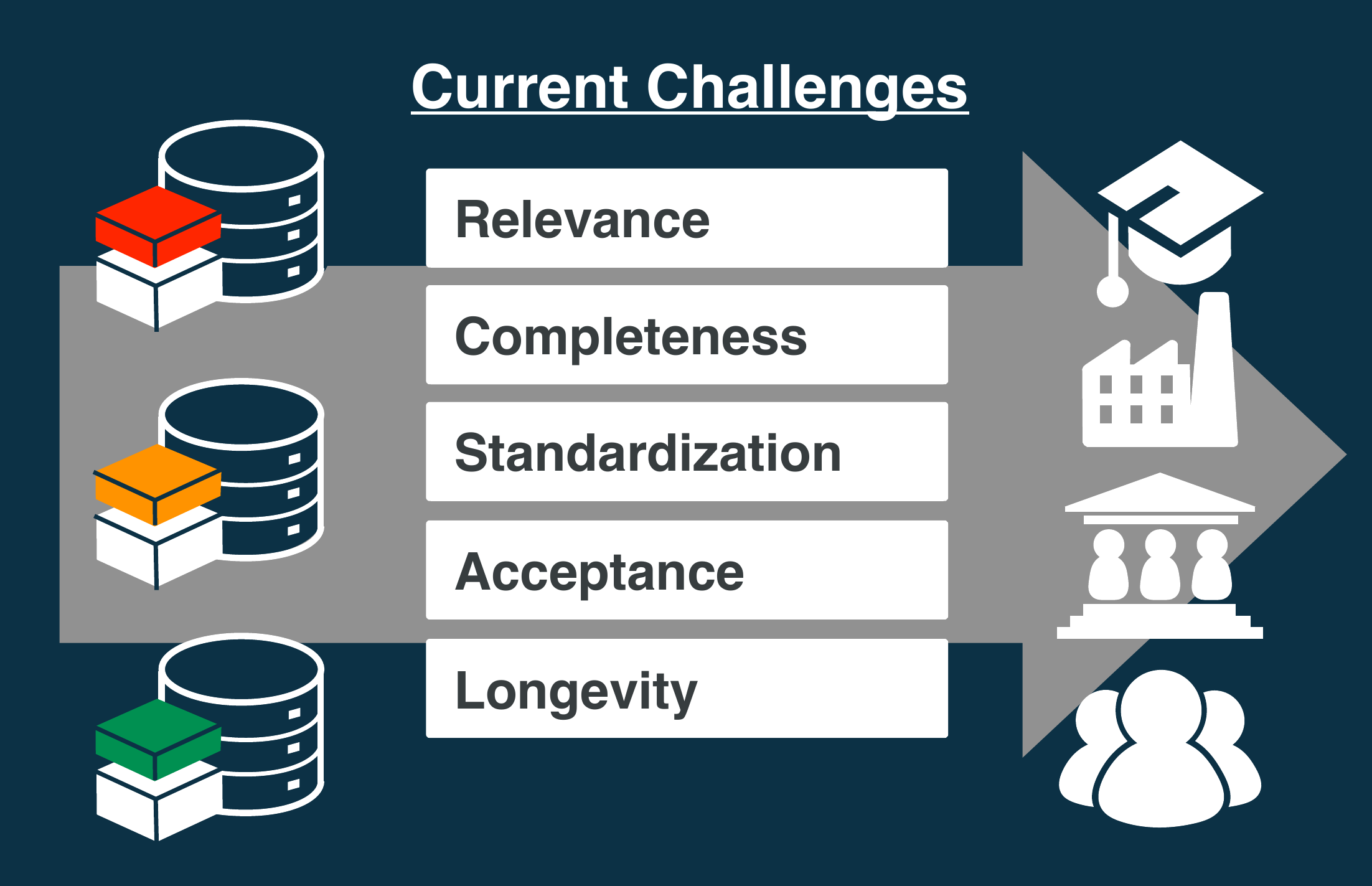}
  \caption{Challenges faced by materials data infrastructures (on the left) on the way to increase the adoption by stakeholders from academia, industry, governments and the public (depicted on the right).}
\label{fig:challenges}
\end{figure}

Having reviewed the current state of data infrastructures in materials science, we now return to the MUSE analogy. The previous two sections illustrated that despite enormous process in data-driven materials science, several challenges need to be overcome before a powerful materials search engine and discovery tool takes shape. The challenges are depicted in Figure~\ref{fig:challenges} and are raised here briefly before being  discussed in  detail in  corresponding sections.

{\bf Relevance and adoption --} Materials data infrastructures must provide relevant data and information to be adopted by stakeholders, be it scientific communities, industries, governments, or the public. Relevance is determined by \emph{data volume}, \emph{data type}, and \emph{data quality}, and entails \emph{data completeness} and \emph{data homogeneity}. Different communities will have different specifications of these terms, which makes it challenging to develop general and interdisciplinary infrastructures that can be adopted. Relevance and adoption are therefore closely related to the subsequent challenges of \emph{completeness}, \emph{standardization}, and \emph{acceptance}.

Relevance also includes tools that operate on the data and help users to classify, analyse, and correlate data. Machine learning has gained the most prominence in this regard and is reviewed in Section~\ref{sec:ML}. Since machine learning is always data hungry, it makes sense to integrate machine learning applications directly into materials data infrastructures. Challenges to such one-stop-shop solutions, which would increase the \emph{acceptance} of materials infrastructures, include the wide variety of available machine learning approaches and data diversity. For data to be informative for machine learning algorithms, its features and properties need to be \emph{relevant} and \emph{complete}.   

{\bf Completeness --} Completeness is ``the quality of being whole or perfect and having nothing missing''. While ideal completeness is hard to attain in practice,  data infrastructures today suffer from a real, severe completeness problem: they contain mostly \emph{computational} and almost no \emph{experimental} data (cf. Table~\ref{tbl:features}).  This state of affairs is rooted in the historic development of data-driven materials science presented in Section~\ref{sec:history_of_time}. Since computational data comes in a digital format, computational scientists were early adopters of database platforms. Moreover computational data is currently more homogeneous and easier to curate than experimental data\cite{roadmap}. Facilitating a seamless comparison between computational and experimental data is, however, an important step toward validating theoretical predictions and in driving materials discovery and development efforts \cite{roadmap}: materials that are identified as promising still require further evaluation, selection, and experimentation. Building synergies among computational and experimental databases thus remains an important challenge for the future of data-driven materials science, which we address in Sections~\ref{sec:ontology} and \ref{sec:creation}.  

{\bf Standardization --} Some form of standardization is essential in the widespread adoption of a new paradigm or technology\cite{Garud/etal:2002,Brunsson/etal:2012}. Stakeholders can only participate in the development of a technology if they speak a \emph{common language}. The language analog in data-driven materials science is \emph{metadata}. Metadata provides relations (the grammar) between data items (the words). Developing standardized metadata for materials science that is  \emph{informative}, \emph{exhaustive}, and \emph{adaptable} is an outstanding challenge. We  address the first steps toward creating a materials \emph{ontology} in Section~\ref{sec:ontology}. Such an ontology, or classification system, would be the foundation for materials science metadata and the evolution of different materials science dialects into a common language. 

{\bf Acceptance and ecosystems --} Materials data infrastructures will only be useful if they are accepted as a useful tool by various stakeholders. Apart from being relevant and complete, data infrastructures have to be user friendly to be widely adopted. User friendliness includes \emph{easy upload} and \emph{download} of data. Easy data upload also facilitates completeness since it reduces hurdles to data sharing. Widespread acceptance furthermore requires trust in the stored data, and this can only be achieved through \emph{data curation}. Data curation is the management and quality control of data throughout its lifecycle, from creation and initial storage to the time when it is archived for posterity or becomes obsolete and is deleted. We address the challenges pertaining to data creation and curation in Sections~\ref{sec:creation} and \ref{sec:quality}.

Infrastructure acceptance is different for different stakeholders. Current infrastructures are predominantly built and used by scientists in academia, as detailed in the previous section. Industry interest and participation has not been systematically studied, and it is dependent mostly on anecdotal evidence. Some materials companies leverage the value of reference databases (e.g., IBM\cite{IBM} and ASM International\cite{ASM}), while others contract the services of intermediaries (cf. Table I and Table II). Apart from this, industry seems to still be exploring the opportunities and potential benefits of materials informatics \cite{Meredig} without full engagement with academia. 

The disconnect between academic and corporate R\&D in many fields makes industry involvement more difficult in this specific case. A hurdle to widespread industry adoption is the \emph{materials gap} - the fact that industry requires other data than what is currently stored in the available materials data platforms. Ecosystems that facilitate the interaction between academic, corporate, governmental, and public stakeholders are a potential solution that we discuss further in Section~\ref{sec:industry_uptake}.

{\bf Longevity and diffusion --} With increasing awareness for open and data-driven science, national and international funding for the development of Open Science (see Sec. \ref{open_science}) is rising, and new materials data platforms are emerging in their wake. However longevity and diffusion of innovations and new technologies are rarely considered by  funding agencies, and long-term financial support for sustained operation is not guaranteed. The initial wave of digital materials infrastructures were built predominantly by materials scientists whose main focus lies in basic science. The long-term maintenance and usability of infrastructure is often only a secondary priority for most scientists. As a result, these digital infrastructures are in danger of becoming digital ruins of the expansion of Open Science. We discuss potential solutions in Section~\ref{sec:industry_uptake}.

Next we explore central topics and applications around data-driven materials science to provide insight into these challenges and into the successes of data-driven materials science.

\section{Materials ontology}
\label{sec:ontology}

We begin our more detailed review sections with the \emph{relevance}, \emph{completeness}, and \emph{standardization} challenges. One of the first decisions in the planning of a materials data infrastructure is which types of materials will be relevant to its intended user base. The most complete representation of these materials will then have to be stored in the database of the infrastructure. The storage  requires standardization and the development of metadata formats. Storing only the raw materials data without any metadata would be futile because raw data is neither searchable nor suitable for machine learning. 

\begin{figure*}[t!]
  \centering
    \includegraphics[width=\textwidth]{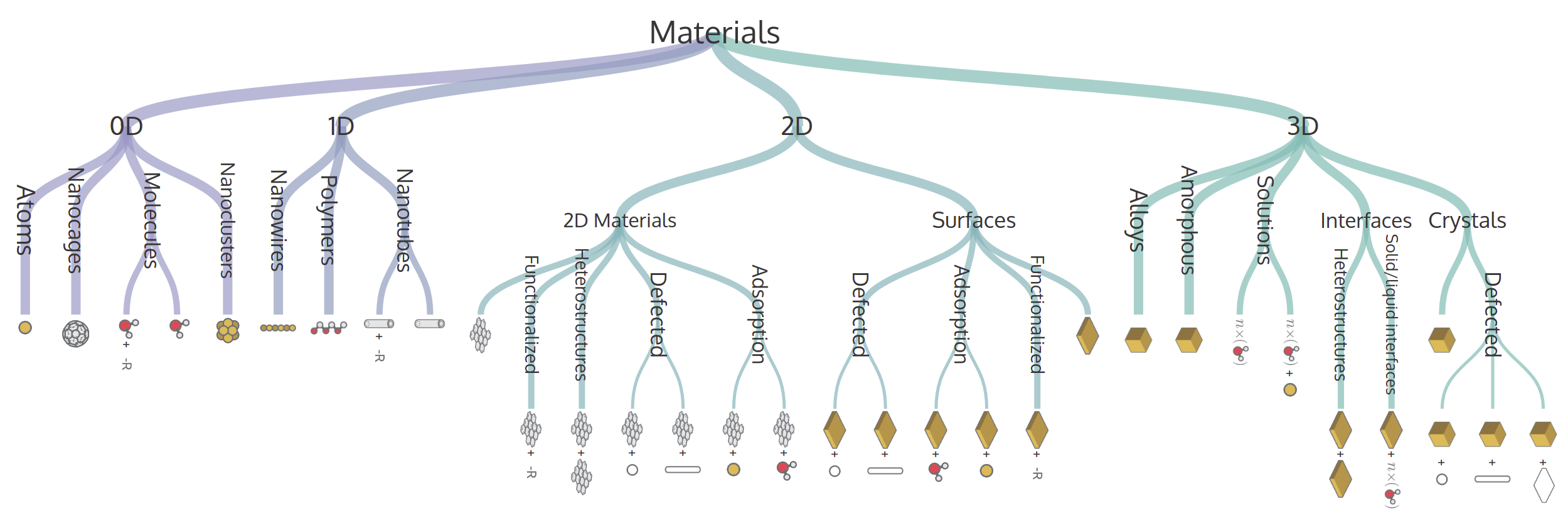}
  \caption{Example of an ontological hierarchy for the structural characterization of materials: a materials tree of life. Figure adapted from \cite{matid}.}
\label{fig:structural_ontology}
\end{figure*}

The development of a metadata framework requires materials classification schemes from which metadata entries and relations can be derived. Crude classification schemes group materials by their functional properties (electronic, optical, mechanical), topological characteristics (bulk, surface, nanotube, polymer, see Fig. \ref{fig:structural_ontology}), or by material type (ceramics, metals, glasses, polymers, or composites). More sophisticated classification schemes are clearly needed to facilitate data-driven materials science. In addition, the origin of the raw data needs to be encoded in the metadata. For real samples, these would be the synthesis and processing conditions and the history of the sample since creation. For virtual samples, the generating computer code and the computational settings need to be known. This clearly demands the classification and organization of materials data in a \emph{materials ontology}.

Ontology is originally a field of philosophy defined as the study of properties, events, processes, and relations of existence\cite{ontology}. In computer science, the term ontology has been co-opted to more specifically mean a formal collection of entities, relationships between those entities, and inference rules that are shared by a community. In materials science, a materials ontology would be a classification scheme for materials, their properties, units, and limits, and their interrelations. Defining an ontology is conceptually important for the purpose of establishing a standard that can be shared by different people working with the same data, and it is a practical necessity in database design. The ontology concept is also closely linked to the ability to search data: the ontology defines the available search terms and facilitates semantic reasoning, which then facilitates complex searches.

Creating the necessary machinery for ontologies in materials science is a tremendous task. An ontology structure has to be developed, suitable formats and standards for encoding meaning have to be defined, and wide-spread adoption of the ontology has to be ensured. For example, a substantial part of information available on the internet today consists of human written text. To interpret the information contained within this text requires human reasoning or sophisticated natural language processing software. But there is a complementary standard by the World Wide Web Consortium called Semantic Web that defines an explicit, machine-readable format (Resource Description Framework, RDF) to organize information on the web. It provides an ontology language (Web Ontology Language, OWL) to describe ontologies for sharing concepts across content creators \cite{semanticweb2}. If this semantic web standard were embraced by the web community, it would significantly boost information sharing across the internet and unleash the power of automated semantic reasoning by artificial intelligence \cite{semanticweb}. As of now, this powerful idea remains largely unrealized.

Similarly in materials science, a standardized ontology that ensures a complete representation of materials has not yet emerged. Currently various ontologies and less-than-formal standards compete. NOMAD Meta-info\cite{nomadrest, metainfohome}, ESCDF\cite{nomadrest}, and OpenKIM\cite{openkim} are the first attempts to categorize computational results in atomistic materials science. PLINIUS\cite{plinius} is used in the field of ceramics, ONTORULE\cite{ontorule} in the steel industry, SLACKS\cite{slacks} for laminated composites, and PIF\cite{PIF}, Ashino\cite{ashino}, EMMO\cite{emmo}, MatOnto\cite{matonto}, Premap\cite{premap}, and MatOWL\cite{matowl} represent general materials science data. Although the development of these materials ontologies has accelerated, they are not nearly as mature as in other fields, for example, the biosciences\cite{ontologysurvey}. Especially for industrial purposes, these publicly available ontologies are typically insufficient, forcing companies to create their own internal, domain-specific ontologies\cite{emmo}.

The lack of standardization aggravates data sharing. Computational science, for example, still relies on file-based data exchanges between different codes. Such file-based data exchange requires interfacing software, significant human resources, and expertise on how the data is structured. Moreover incompatible standards lead to conversion errors and data loss. These interoperability problems could be solved by a common ontology and a standardized representation of knowledge within this ontology. Fitting existing and novel data into such an ontological framework would still be a tedious and error-prone task for humans. Existing tools and techniques could, however, be used to simplify and automate this process. Data curation services\cite{qresp} help in organizing and annotating data, natural language processing can be used to mine data from scientific literature\cite{nlp1, nlp2}, and automated structural classification helps in categorizing the contents \cite{matid, spglib, aflowsym}.

The ontologies themselves could also be constructed semi-automatically by observing the nomenclature and the relation of concepts used in the literature\cite{ontologylearning1, ontologylearning2}. If widely adopted in materials infrastructures, this standardization would enable the vision of a powerful search platform for materials science. Once defined and filled, AI solutions would benefit from it. One could envision virtual AI agents helping scientists to answer complex questions related to material performance and synthesis by analyzing materials databases and scientific literature. Such AI agents would not only aid basic research, they would also help businesses that could then more effectively leverage existing scientific knowledge in their R\&D.

Recently there have been promising efforts in trying to unify the nomenclature and standards in materials science by the European Materials Modelling Council \cite{emmo}, the Research Data Alliance (RDA)\cite{rda}, and by a collaboration between NOMAD scientists and the Centre Europ\'een de Calcul Atomique et Mol\'eculaire (CECAM) \cite{nomadrest}. A concrete example of such collaboration is the Open Databases Integration for Materials Design (OPTiMaDe) consortium\cite{optimadeurl}. OPTiMaDe is building a common interface for accessing data from multiple materials platforms. The diversity of subfields and stakeholders in materials science might make it impossible to define one universal materials ontology. We, however, recommend that unifying ontologies whenever possible and disseminating these efforts to the materials science community are key steps in making the most out of the rich body of materials data created with the modern experimental and computational methods discussed next.

In summary, standardization facilitates data sharing. A materials ontology is a classification scheme for materials that enables standardization. Ontologies also ensure completeness of materials data since everything that falls outside of an ontology by definition indicates a lack of completeness in the ontology. Attempts at constructing materials ontologies are underway. However more needs to be done to ensure that relevant materials and relevant materials properties are incorporated into existing materials infrastructures. Otherwise our MUSE will return irrelevant information when queried.

\section{Data creation}
\label{sec:creation}

We now stay with the challenges of relevance and completeness and address how enough relevant data can be generated to feed a materials infrastructure. Once again, we encounter standardization but this time in the context of standards for generating data. We briefly review techniques and recent improvements in data creation methodology - so-called high-throughput methods - that are enabling  experimental and computational scientists to efficiently create data for data-hungry repositories and applications.

For experimental materials data, the introduction and refinement of deposition and analysis methods has had perhaps the greatest impact on data creation efficiency. In 1965, the first composition gradients could be achieved in thin-film material co-deposition\cite{hte}, offering a more efficient replacement for the one-by-one creation and study of materials. Since then, multiple improved materials synthesis and characterization techniques have been introduced.\cite{hte2, hte3, clmbe, Fukumura/etal:2000, Kneiss/etal:2018} They have enabled the rapid generation of composition-structure-property relationships \cite{Koinuma/Takeuchi:2004, Krishna:2008, Potyrailo/etal:2011, Chikyo:2011, vonWenckstern/etal:2015}. State-of-the-art deposition techniques, such as combinatorial laser-molecular beam epitaxy (CLMBE)\cite{clmbe} introduced around the year 2000, can be used to create temperature and composition gradients across the sample and provide control of the deposition in three dimensions. 

These new methods facilitate a finer and more complete sampling of structural phases and chemical compositions in a single experiment. They efficiently create \emph{materials libraries} -- experimental samples with one or more composition or phase gradients. Each library represents part of a well-defined materials space. Measurements from such materials libraries are now made accessible through Open Access online services, such as the High Throughput Experimental Materials (HTEM) database\cite{htem}.

In contrast, and perhaps surprisingly, the high-throughput creation of computational materials data has only become common practice in the 21st century\cite{aflow, oqmd,highway}. Thanks to Moore's law and massively parallel computing architectures, available computational power has increased steadily, and computational data creation has quickly taken advantage of this power, even surpassing experimental efforts. For example, the Open Quantum Materials Database (OQMD) performs virtual high-throughput materials synthesis by decorating known crystal structure prototypes with new elements. It has now grown from the initial set of roughly 30,000 experimentally known crystal structures from the Inorganic Crystal Structure Database (ICSD) to more than 560,000 computationally predicted materials\cite{oqmd, oqmdurl}. High-throughput workflows have now matured and are increasingly applied to screen also complex properties such as coupling and reorganization energies in organic crystals \cite{Schober2016JPCL,Kunkel/etal:2019}.

In the creation of such massive datasets, it is increasingly important to adhere to computational standards. This standardization has been pioneered by the Materials Project and the AFLOW-consortium (see Table \ref{tbl:databases}), with comprehensive specifications for the methodological details, such as k-point grid densities, cutoff energies, pseudopotentials, and convergence criteria, related to Density Functional Theory (DFT) calculations \cite{mpstandard, aflowstandard, aflowbandstandard}. This standardization ensures that data can be made cross-compatible within a database or even across databases.

\begin{figure}[h!]
  \centering
    \includegraphics[width=0.48\textwidth]{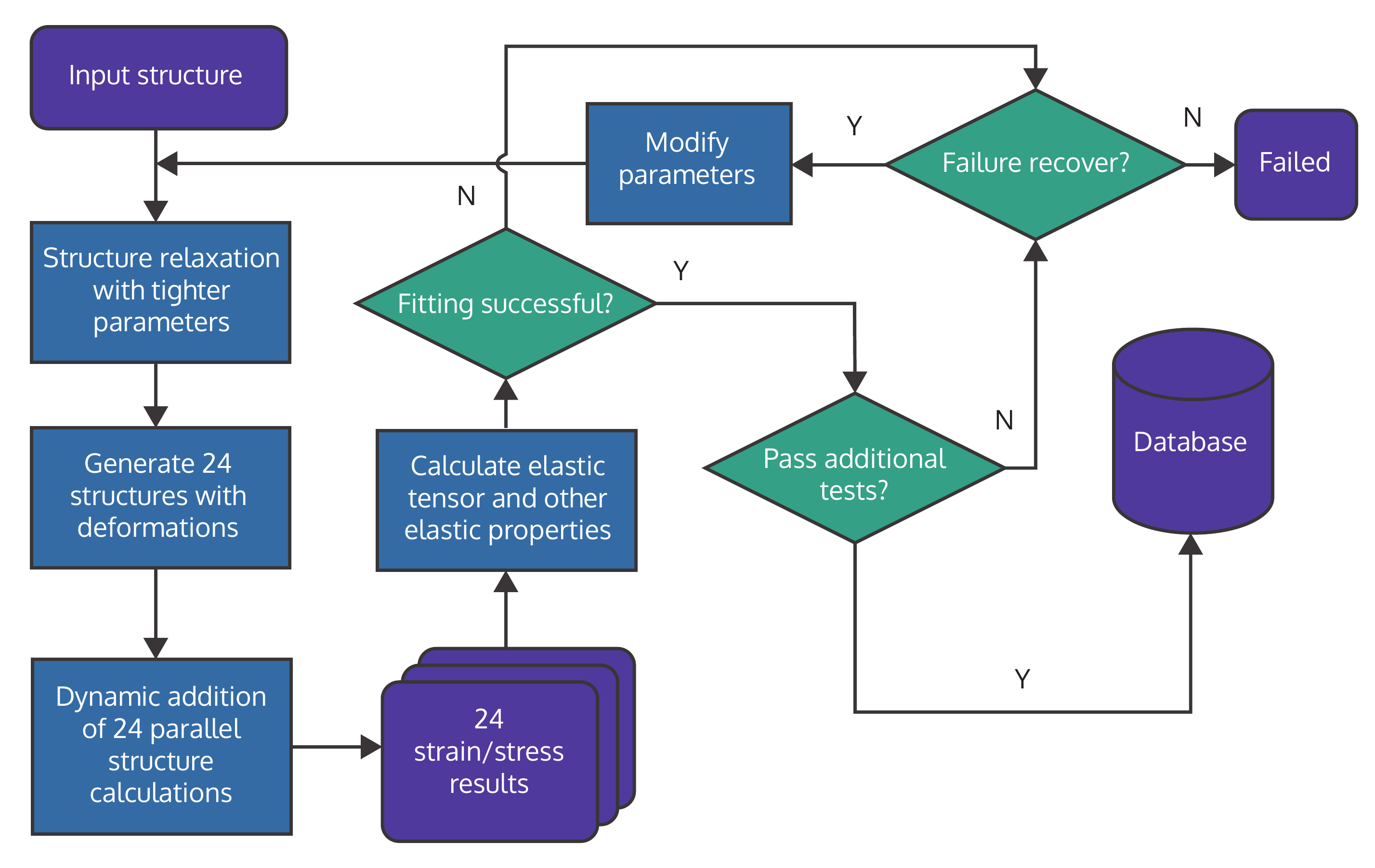}
  \caption{A computational workflow used in creating a dataset of elastic tensors with the FireWorks workflow manager. The indigo boxes correspond to inputs or results, lighter blue boxes correspond to actions, and green diamonds correspond to decisions. Figure adapted from Ref.~\onlinecite{fireworks}.}
\label{fig:workflow}
\end{figure}

Relatively recent additions to computational materials science are workflow management tools like FireWorks\cite{fireworks}, atomate\cite{atomate}, AiiDa\cite{aiida}, and AFLOW$\pi$ \cite{aflowpi}. These tools enable researchers to build automated and robust workflows for creating consistent data sets. Workflows connect different computational \textit{steps} and \textit{checks} into a single computational graph. The computational steps generate data, and checks aid with automated recovery from errors that might occur in a computational step, for instance due to incorrect computational settings or hardware failures. An example of a workflow graph from FireWorks is given in Figure \ref{fig:workflow}. 

In summary, materials data needs to be created in sufficient volumes for materials infrastructures to be relevant and complete enough. To fulfill this need, high-throughput experimental and computational methods have emerged. The level of automation and efficiency provided by these methods ensures that the bandwidth at which materials data can be created should not be an issue for the MUSE of the future.

\section{Data quality}
\label{sec:quality}

From data generation, we move on to data quality. As already alluded to in the previous section, the quality of data is related to \emph{standardization} in the generation of data and considerably affects the \emph{acceptance} of data infrastructures.

Big data is often characterized in terms of four Vs: \emph{volume}, \emph{velocity}, \emph{variety}, and \emph{veracity}. Each V poses a challenge, although the volume challenge could, in principle, be solved by more storage space, and the velocity challenge of faster data generation could be addressed by faster computer processing and accelerated measurement and fabrication techniques. Increased variety is more a challenge for standardization and ontology integration, yet it is also a benefit for machine learning and materials discovery algorithms. Veracity, however, is the most problematic because it is a softer measure of how to quantify the degree of trust in data and how to improve its trustworthiness.

Data veracity has two aspects: bias and variance. In an experiment or a calculation, the bias of the result is quantified as the offset of the average result from the ground truth, whereas variance is quantified by its probabilistic definition as the spread of the results over identical runs. Note that the variance and bias discussed here originate from approximations in theoretical models or experimental uncertainties, not from stochastic processes in the experiment or calculations, for example in molecular dynamics simulations. We also disambiguate the use of bias and variance here from the same terminology commonly used in machine learning.

For users of data infrastructures, it is important to know the quality of available data sets. However both data bias and its veracity may be hard to quantify.

In the computational realm, variance can be caused by different computational environments or differences in implementation but it is typically negligible even between different software implementations\cite{dftreprod}. Computational variance is also often one or several orders of magnitude lower than the corresponding experimental variance\cite{elementalcrystals}. The veracity of computational data is thus dominated by bias -- offset from the experimental ground truth. The estimation of this bias depends on access to experimental data or comparison to results from a higher-level theory \cite{mse}. The bias also depends on the types of chemical elements in the materials. Some computational approaches or approximations  may break down for specific groups of elements, such as dispersion-governed compounds, magnetic materials, strongly correlated materials, and relativistic effects in heavy elements, leading to much larger errors for these groups of materials\cite{elementalcrystals}.

While computational scientists have full control over their simulations, experimental scientists often face errors that are beyond the control of their experimental setup. Bias in measurements can be due to incomplete knowledge of a sample's content and history, as well as interactions between the sample and the environment. Variance can be caused by material imperfections and contaminants, and experimental uncertainties introduced by the equipment. As such, it is typically difficult to discriminate between bias and variance in experimental errors. If there are no comparable experimental facilities, this can also make it hard to assess the data quality. In some cases, quality-controlled commercial equipment and widely accepted standard procedures are available when performing measurements. However, there is often a need to use custom-built equipment or to measure materials for which the standard procedures are not applicable, making it harder to reproduce and validate results. This difficulty of controlling more elaborate experimental setups means that reliable experimental data exists for simple systems, such as small molecules\cite{g99} or elemental crystals\cite{elementalcrystals}, but for more complex systems and measurements, the data quality may be harder to determine.

The combination of high bias in computational results and the difficulty of controlling errors in experiments makes the overall estimation of data veracity in materials science particularly hard. One example is given by the formation energies of crystals for which computational and experimental values notably differ. This discrepancy is caused by both systematic errors in the computational methodology and experimental uncertainties\cite{Kirklin2015}.
Due to the species-specific nature of the computational error and the vastness of compositional and structural space, it is impossible to make an exhaustive brute-force comparison between experiment and computation. However intelligent error extrapolation schemes are being developed. In one such scheme, the computational error of non-converged energies for crystals with two different chemical species  - compared to fully converged energies -  can be estimated by using a linear combination of errors from solids with only one chemical species\cite{nomaderror}. Such schemes could also be extended to estimate the error between experimental and computational data. This will require systematic data collection from both experiment and computation but it may prove to be fertile for practical error estimation.

In summary, data quality is important to ensure standardization of data and to increase acceptance of data infrastructures, but it is challenging to quantify. Two indicators of quality are bias and variance. Systematic data collection and new extrapolation schemes would facilitate bias and variance assessments in the future.

\section{Data analytics}
\label{sec:ML}

To increase the \emph{adoption} of data infrastructures, developers are adding tools and apps to their data platforms that operate directly on the data in the infrastructure's database. These tools add value to the data and enhance its \emph{relevance}. Many tools now include machine learning. Here we briefly review the main types of machine learning and illustrate how they could add value to materials infrastructures.

\subsection{Introduction to machine learning}
Machine learning is the scientific study of how to construct computer programs that automatically improve with experience\cite{mitchell1997machine}. More specifically, machine learning algorithms use statistical models and optimization algorithms to reveal patterns in training data  to make predictions or decisions without being explicitly programmed to perform a certain task. The advantage over human learning is that computers can often handle much larger and higher dimensional data, and suitable approximations can be automatically found by monitoring how well the models generalize to unseen data.

Many of the statistical methods in machine learning have been around for decades. For example, Marvin Minsky built the first hardware implementation of a neural network\cite{minsky} in 1951. Our current AI boom has been facilitated by the rapid hardware development for information storage and processing, the conscious effort of data gathering and curation, as well as increased developments of machine learning methods and libraries by private companies, the public sector and academia, driven by the potential that machine learning can unlock from previously untapped data resources. Today machine learning is a key ingredient of materials informatics, as showcased by various reviews on the topic\cite{RAJAN200538, Rajan:2015kf, Agrawala, Bartok/etal:2017, Liu2017, Gubernatis:2018gi, Ramprasad2017, Ward2017, Butler2018, C9TA02356A}.

Machine learning can be divided into different subfields that are characterized by the available data. \textit{Supervised learning} is the most mature and powerful of these subfields and is used in the majority of machine learning studies in the physical sciences\cite{Butler2018}. Supervised learning applies in situations where a machine learning model is trained on input-output pairs from a real process to produce optimal outputs for unseen inputs. Typical applications are predictions of physical properties (like formation energies\cite{Isayev2017, cgcnn, voronoi} or molecular properties\cite{cm, schnet, alchemy, Gosh/etal:2019, Annika_KRR}) given the input features of a material or process (e.g., geometry, physical properties, external conditions). 

In \textit{unsupervised learning}, only input data is given to a model but no output. The machine is then tasked with a learning objective, for example to find rankings or patterns for this input. Unsupervised learning is often used to pre-process input data, such as dimensionality reduction by principal component analysis (PCA)\cite{jolliffe2002principal, pcaapplications}, or aiding the analysis of complex output data like visualization of high-dimensional data with T-distributed stochastic neighbor embedding (T-SNE)\cite{tsne, htem} or sketchmap\cite{sketchmap}. 

Finally \textit{reinforcement learning} is a rapidly emerging field with promising applications in tasks that require machine creativity. In reinforcement learning, a model is given a task of choosing a set of actions to optimize a long-term goal. As such, it differs from supervised learning because no correct input-output pairs are presented for individual actions, but the training is a mixture of exploration and exploitation guided by a long-term reward \cite{bishop2006pattern}. This mode of learning can be useful in the exploration of compound and material spaces like exploration of grain-boundary structures with evolutionary algorithms\cite{grainboundaries} and the search for new molecules with objective-reinforced generative adversarial networks\cite{Guimaraes2017}. 

\begin{figure}[h!]
  \centering
    \includegraphics[width=0.45\textwidth, ]{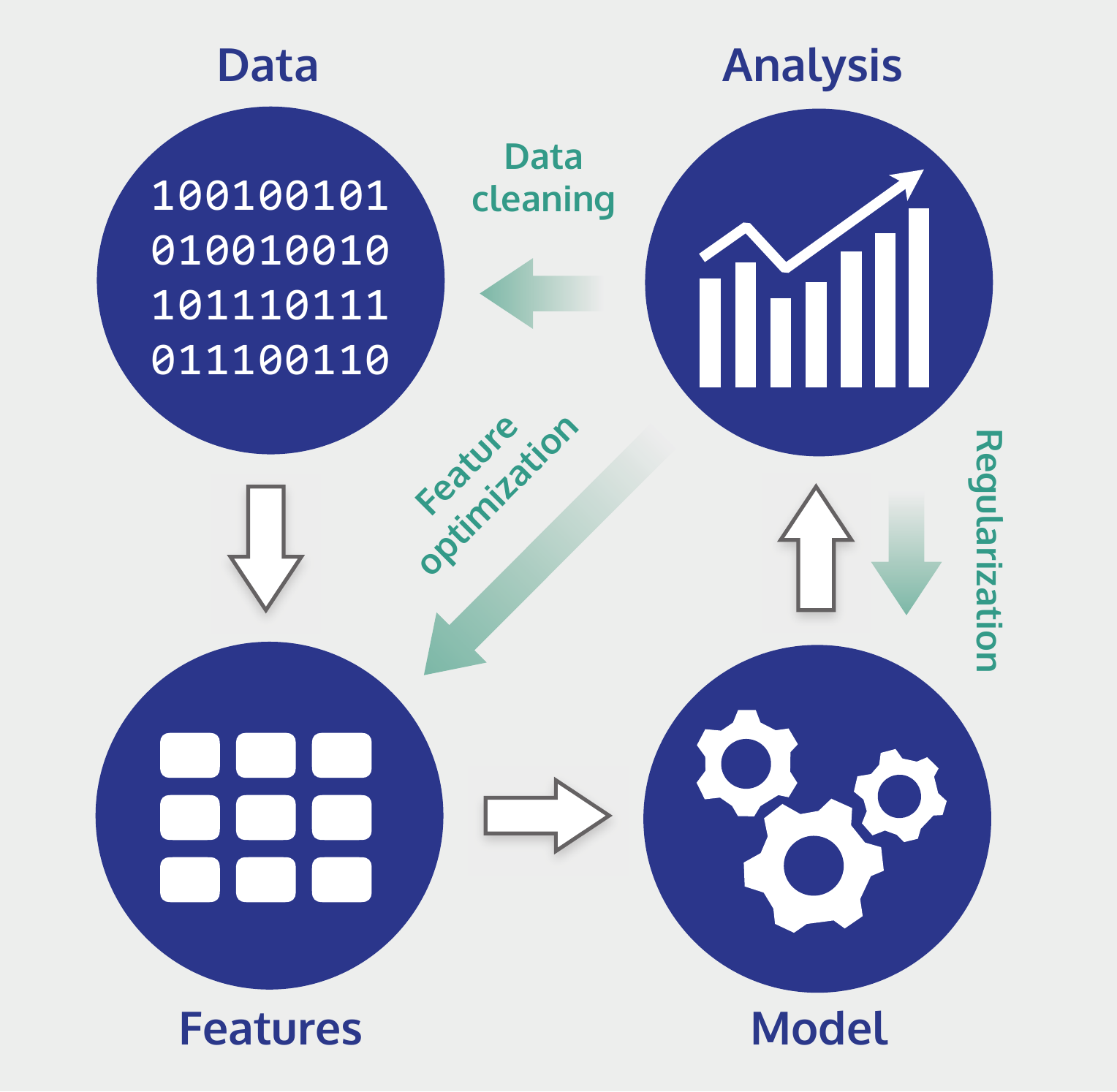}
  \caption{Key steps in building a machine learning model. The white arrows indicate the flow of data, green arrows indicate actions that can be identified and performed after analysis to improve the performance of the model. }
\label{fig:ml_workflow}
\end{figure}

The knowledge contained in a machine learning model is encoded in the parameters of the model, and it is, in principle, tractable. However the number of parameters can reach into the millions, which makes these models quite opaque to human interpretation. This is different from the scientific approach thus far, which has relied on deriving and discovering physical laws that are encoded in humanly readable equations. For commercial applications, the transparency of the models is not as important as their performance, but for advancing scientific understanding and wider acceptance, better human interpretability would be beneficial. Recent examples of approaches that analyse machine learning models to reveal their mechanisms include the analysis of input feature importance\cite{Isayev2017}, explicit formulation of the input in algebraic form\cite{Ghiringhelli:2015kra, sisso}, and analysis of convolutional neural network filters\cite{Ziletti2018}.

\subsection{Specifics of machine learning in materials science}
Currently the applications of machine learning in materials science are rich and diverse, ranging from catalyst design\cite{Goldsmith:2018bb, marcaki}, exploring the mechanisms of high-temperature superconductivity\cite{cartography, Stanev2018}, to predicting excitation spectra\cite{Gosh/etal:2019}. Building such applications can generally be broken down to four key steps: data acquisition, feature engineering, model building, and analysis. These steps are illustrated in Fig. \ref{fig:ml_workflow}. They are, however, interdependent and often multiple iterations of each step are required to create a successful machine learning system. Specialized software frameworks\cite{matminer, chemml, deepchem} have been developed to aid the set-up and build and management of machine learning models.

\begin{figure}[h!]
\begin{overpic}[width=0.48\textwidth, scale=1.0,unit=1mm]{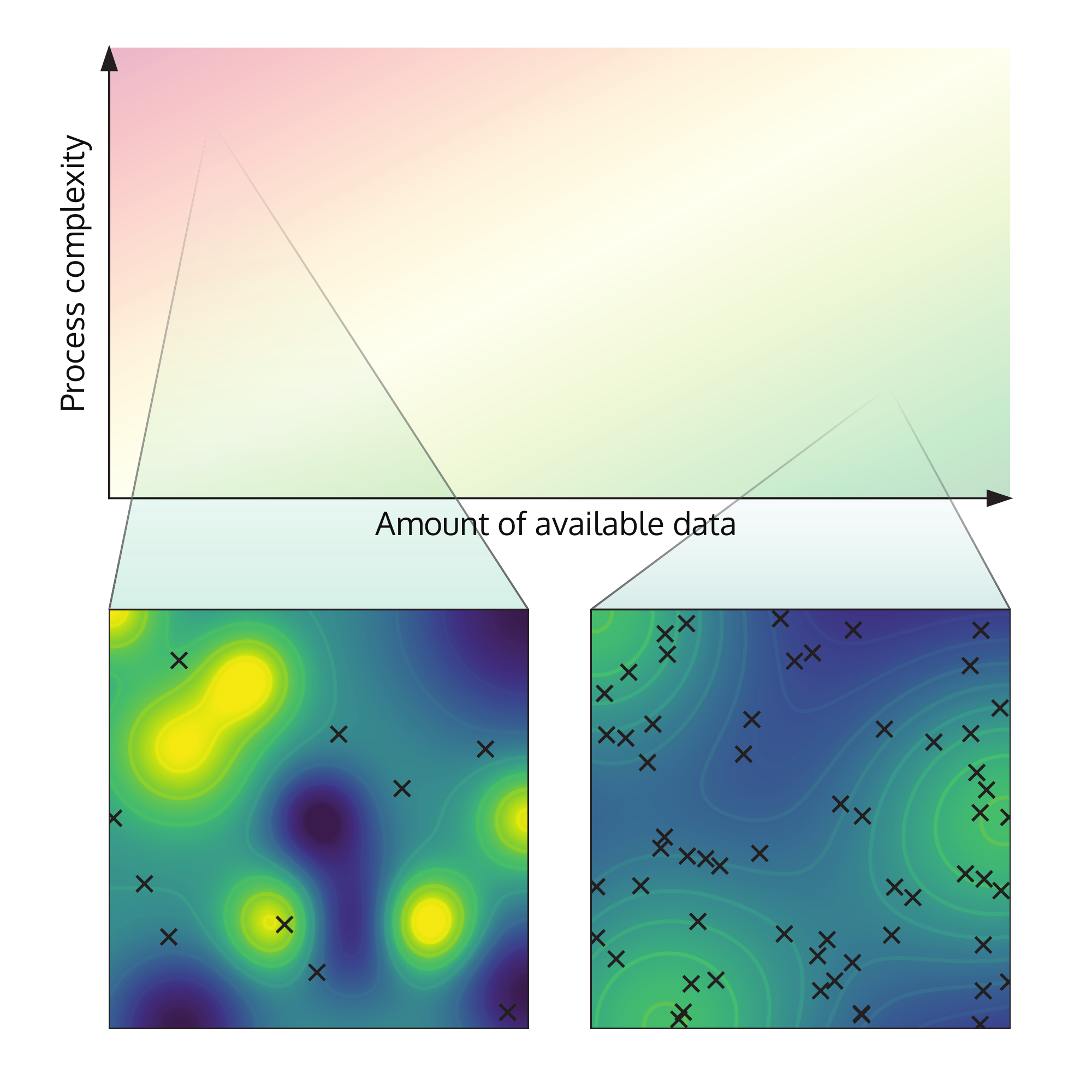}
    \put(25,67){\parbox{2.5cm}{\tiny{Band gaps\cite{cmr_application, applicationpv3, cgcnn, Isayev2017}}}}
    \put(50,61){\parbox{2.5cm}{\tiny{Forces for molecular dynamics\cite{ani1, gap, schnet, onthefly}}}}
    \put(42,53){\parbox{2.5cm}{\tiny{Molecule energies\cite{cm, schnet, soapapplications, alchemy}}}}
    \put(12,75){\parbox{2.5cm}{\tiny{Microstructure optimization and prediction\cite{microstructure}}}}
\end{overpic}
\caption{The machine learning domain in terms of data volume and the complexity of the physical process, with selected examples placed in this domain. The complexity of a physical process here means the complex, nonlinear structures present in the data. Two opposing learning scenarios, a hard and an easy one, are illustrated in the lower panel. In these two cases, the underlying physical process is represented by a coloured contour map, and the sampling of this process is represented by black crosses.}
\label{fig:ml_learning_map}
\end{figure}
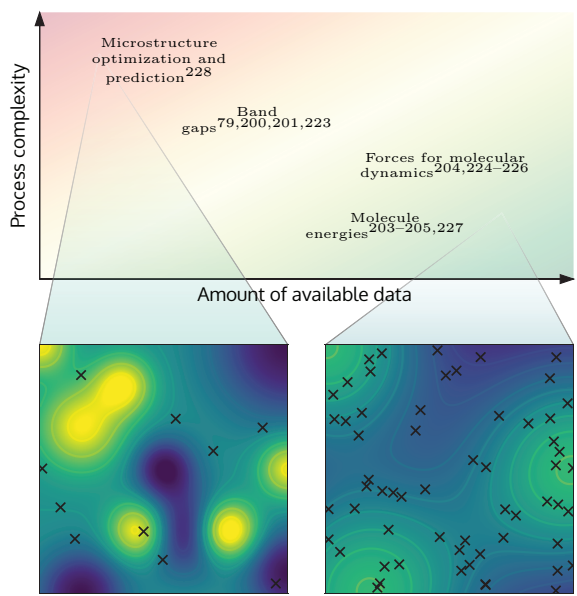

While machine learning generally requires data, the amount of data depends on the specific problem. Figure \ref{fig:ml_learning_map} illustrates the trade-off between the available data volume and the complexity of the underlying process for different machine learning approaches. Problems in the top-left corner are not suitable for machine learning due to the low amount of available data. The further to the right a problem sits, the more suitable it becomes for machine learning. In practice, it is often difficult to place machine learning methods and new problems in this diagram. Thus rapid prototyping of the problem is frequently key to successful machine learning. Since we have control over only one of these parameters - the amount of data - the importance of open data access, materials databases, efficient data creation, and data veracity is paramount for the success of machine learning.

Machine learning models expect input data in alpha-numerical form, typically as an array of letters or more often as numbers. Raw data, however, is usually unsuitable for machine learning. The first task in building a machine learning model is therefore to extract informative features from the raw data (cf. Fig.~\ref{fig:ml_workflow}). \emph{Feature engineering} refers to the act of introducing domain expertise to the learning model by affecting which features are used. It can be beneficial to apply problem-specific feature transformations, called feature extraction\cite{bishop2006pattern}, which exploit known symmetries in the input for example, making it easier for the model to learn a unique mapping. This can be especially important for input features that encode atomic geometries. Physical properties exhibit symmetries with respect to translation, rotation, and index permutation in the Cartesian coordinates representing a geometry. Using a transformation that makes the input invariant with respect to these symmetries will help the learning process by creating a unique mapping from an atomic geometry to its properties. Such structural descriptors have been successfully applied in the prediction of molecular and crystalline properties\cite{cm, fragment, alchemy}, and there their development has exploded in recent years \cite{cm, cm_versions, bob, sm, soap, mbtr, acsf, wacsf, voronoi, fragment, wyckoff, alchemy, f2bf3b}. To facilitate easier navigation through descriptor choices, application-neutral software libraries for descriptors are being developed\cite{dscribe, qml}. In contrast to human-driven feature engineering, the optimal features can also be discovered more systematically by learning them directly from the data with \emph{feature learning}. In the simplest form this can be achieved by methods like principal component analysis (PCA)\cite{jolliffe2002principal}, which is based on the variance of the input features. In the other extreme the features may be formed by an encoder as a nonlinear latent space within an autoencoder neural network\cite{Goodfellow-et-al-2016}. Analysing the features used by the machine learning model in making a decision forms the basis of understanding and verifying the correctness of the model. Integrating such analysis into the workflow of building machine learning models is still lacking in many cases, hindering their acceptance and interpretability.

After feature selection (cf. Fig.~\ref{fig:ml_workflow}), the machine learning algorithm must be chosen (see different machine learning types discussed at the beginning of this section). Each algorithm has its own application domain, and there is currently no algorithm that is optimal for all problems \cite{Liu2017}. This conundrum is also known as the ``no free lunch theorem''\cite{nflt}. Some common choices include feed-forward neural networks\cite{nn}, decision trees\cite{dt}, kernel ridge regression\cite{kernel}, support vector regression\cite{kernel}, and Gaussian processes\cite{gp}. These approaches are common in computer science and are available in generic software packages that help select the best model for a task\cite{scikit-learn, keras, tf, xgboost, caffe, theano, pytorch, gpy, DeepLearning4J}. Apart from such generic approaches and packages, machine learning models are often customized to materials science. One example is the creation of custom neural network architectures\cite{cgcnn, dtnn, mpnn, schnet,Gosh/etal:2019} that have been designed specifically for atomistic geometries, reducing the need for feature engineering.

One practical concern in model selection is the amount of data that is available for training the model. Methods like kernel ridge regression require an inversion of a matrix whose size is proportional to the number of training samples. This restricts their usability for large datasets because the time taken by a brute force inversion scales as $\mathcal{O}(n^3)$ with the dataset size $n$. Other models like neural networks can handle larger datasets since they can be trained by using small batches of the training data, and their performance can be monitored during training. At the other end of the spectrum we find powerful tools for small datasets such as regression with Gaussian processes and Bayesian optimization\cite{Milica_BOSS, Gubernatis:2018gi}, the extraction of effective materials descriptors with subgroup discovery\cite{subgroupdiscovery} or compressed sensing as done in, e.g., the least absolute shrinkage and selection operator (LASSO) or the sure independence screening and sparsifying operator (SISSO) \cite{Ghiringhelli:2015kra, sisso}. Also some forms of input are better suited for certain models. For example, images exhibit a high degree of correlation between adjacent input points. Models that exploit such correlations, like convolutional neural networks\cite{LeCun}, may then be the best choice. In other cases, the input features have no apparent correlations or have completely different numerical ranges, and decision trees may exhibit the best performance\cite{dt}.

The final step in the machine learning workflow is the performance assessment (cf. Fig.~\ref{fig:ml_workflow}). This analysis guides all other aspects of learning - from excluding corrupt learning samples to optimizing the features and the model itself. The analysis step is general for all application areas of machine learning. For a more in-depth introduction, we refer the reader to existing literature\cite{bishop2006pattern, mitchell1997machine}. The goal of this step is to ensure that the model generalizes well to unseen data. Two common problems are \emph{over-} and \emph{under-fitting}. In over-fitting, the model becomes too specific. It reproduces the training data very well but performs poorly on new data. In under-fitting, the model learns rules that are too general and averages through training and through new data. The balancing act between over- and under-fitting is called the bias-variance trade-off, and it is typically controlled with cross-validation and careful data set design. The whole data set is usually split into a training set, from which a further validation set for hyper-parameter optimization can be split off, and a test set. Model performance is then evaluated on the test set. The dataset contents and the exact way the data is split into training and test sets can affect the reported performance and in some cases lead to unrealistic results. Often the sampling of training examples is not very even in the input space, as the samples can exhibit high levels of clustering -- for example the dataset may comprise of multiple clustered material types that have very similar properties. In such cases the model is able to interpolate very well even if it has only been trained on one representative of each cluster, but its performance will start to deteriorate for unseen material types, which are hard to leave out of the training set with purely random selection. Due to this effect randomly split training and test sets can offer unrealistic performance metrics and alternative cross-validation strategies like leave-one-cluster-out cross-validation (LOCO CV)\cite{lococv} offer more realistic performance metrics.

All the key elements for successfully applying machine learning in materials science are in place, as illustrated also by the applications showcased in the next section. However, several challenges prevail. For example, selecting the optimal combination of data, features, machine learning models and analysis tools can be a formidable task, especially because the field is advancing so rapidly and practices become outdated quickly. Careful curation and standardization of both data and machine learning models can to some degree mitigate the problem, but not enough benchmark sets have been established in the community. Also,  available data volumes are often still too small to apply machine learning tools that have been successful in other domains, e.g., commerce or social media.

Another challenge is the exchange of pre-trained models. Projects such as OpenML\cite{OpenML2013} and DLHub\cite{DLHub} are first examples for model-and-data sharing platforms that enable transfer learning, but more could be done. Metric assessment is a further challenge. The reported performance for machine learning models is an important selection criterion for adopting certain models or features.  However, performance metrics are not yet standardized. Standardized datasets help, but more attention should be devoted to the selection of test and training sets to obtain more realistic error bars. 

We have already discussed the challenge of interpretability. As the exact way input data informs the machine learning model is often blindly guided by the model optimization and hidden behind internal parameters, better methods for interpreting the decisions made by machine learning models are required. Although the natural sciences rarely have to worry about  ethical consequences -- unlike the social sciences that are now adopting AI into their decision making \cite{irving2019ai} -- a critical evaluation of the decision mechanisms is important for understanding the shortcomings of machine learning models and to advance scientific understanding.

In summary, machine learning is a powerful concept for data analysis and materials informatics. Machine learning is a field undergoing very active development, and a plethora of suitable machine learning methods has been applied to materials science. Increasingly such machine learning tools are incorporated directly into data infrastructures. In our MUSE analogy, they will provide meaningful answers to our ``searches''.

\section{Applications}

Staying with \emph{relevance} and \emph{adoption}, we now briefly present areas in which data-driven materials science has been applied successfully. Success stories are important for the development of any field as they inspire trust and commitment in stakeholders. 
We have identified three major research objectives for which we think data-driven approaches have the largest impact on materials science: \emph{materials discovery}, \emph{understanding materials phenomena}, and \emph{advancing materials modelling}. We review these three areas briefly and present relevant studies.

\emph{Materials discovery} is a complex problem in which a list of target specifications are given and the optimal material is sought. Such discovery is often performed by using a forward solution - simply calculating the key properties for a pool of candidate materials to identify the best ones for further in-depth analysis, characterization, and verification. By using existing or dynamically generated materials data, scientists can build heuristic models (often through machine learning) to dramatically speed up the identification of best candidate materials for experimental synthesis. Although experimental synthesizability is often a bottleneck, there are multiple studies that have verified that this approach has the power to accelerate the discovery of novel or improved materials. The examples that have already resulted in experimentally synthesized novel compounds include new molecules for efficient organic light emitting diodes (OLEDs)\cite{oled}, polymer dielectrics for electrostatic energy storage\cite{ramprasadpolymer}, novel gallide Heusler structures\cite{heusler}, NiTi-based shape memory alloys with small thermal dissipation\cite{Xue2016}, lead-free piezoelectrics\cite{piezo1} and metallic glasses\cite{mg} and high-entropy alloys\cite{HEA} for structural applications requiring hardness and corrosion resistance. Furthermore, multiple novel materials identified by this virtual screening are awaiting experimental validation, including photovoltaics\cite{applicationpv1, applicationpv2}, photoelectrochemical water splitting materials\cite{splitting1, splitting2}, topological insulators\cite{TI1, TI2} and novel binary or ternary crystal structures\cite{ternarypred, Meredig2014}. 

If the desired candidate material is not in the pool of materials that are being screened with the forward solution, then it cannot be found in this way. In such cases, the problem has to be inverted, that is, a mapping from the target property space to materials space has to be found. For solid clusters, simple inverse relations have recently been established between X-ray absorption (XAS) spectroscopy \cite{Timoshenko/etal:2017,Timoshenko/etal:2018,Zheng/etal:2018} and coordination shells of atoms. For molecules, neural-network-based auto encoders and decoders \cite{Gomez-Bombarelli/etal:2018} were combined with a grammar-based variational autoencoder \cite{Kusner/etal:2017_arxiv} to map from the discrete molecular space into a continuous latent space (in which optimizations can be performed) and back again. Even with such sophisticated models, it is not easy to generate valid, synthesizeable molecules and materials with the wanted properties, and inverse predictions remain difficult in practice.

Historically new materials were often discovered by first understanding fundamental materials phenomena and then applying them to look for other materials that might fit the same physical laws. Machine learning based predictions conceal these physical laws and do not provide the same understanding of materials-property relations. By changing the objective to \emph{understanding the underlying materials phenomena} (asking why instead of what), we can hope to use reductionistic approaches to transcribe data into laws and equations that are more typically associated with scientific progress. In vast and high-dimensional materials data landscapes, for which human-intuition is ill-suited, this discovery of materials phenomena can be aided by a data-driven approach. The fundamental mathematical formulation of physical laws, such as conservation laws and differential equations, can be automatically deduced from data\cite{Schmidt81, Rudye1602614, RAISSI2019686}. Methods based on compressed sensing\cite{Ghiringhelli:2015kra, sisso} provide a systematic way of identifying the algebraic form of the descriptors that capture the underlying mechanisms behind material properties, providing a more natural basis for human interpretation. These methods have been successful in identifying physically meaningful descriptors that control the stability of perovskites\cite{sissoappl} and monolayer metal oxides coatings\cite{lassoappl}. Another idea is to map high-dimensional data into more easily human analyzable two- or three-dimensional maps with unsupervised learning. This idea of ``materials cartography'' has been used to identify common features for high-temperature superconductor materials\cite{cartography}, to group molecules into intuitive maps that can reveal key structure-property relations\cite{sketchmap}, find phase transitions in complex systems \cite{Wang:2016fu}, or establish the key descriptors in the catalytic properties of metal surfaces \cite{Chowdhury:2018ck}.

The final application area is related to \emph{advancing materials modelling} with automated construction of surrogate models directly from data. These surrogate models can replace the laborious fitting of semi-empirical models, and if trained with highly accurate data are able to reproduce complex chemical phenomena with very low computational cost by sacrificing some of the accuracy. Such AI-based modelling tools are able to assist even in very challenging tasks, as demonstrated by IBM RXN\cite{rxn} -- a free online tool that uses a machine translation inspired architecture to predict the product of chemical reaction from the structural formulas of the reactants. Another example is the creation of classical force fields from DFT training data\cite{gap, ani1, agniuniversal, schnet, onthefly}. 

The promise of these methods is to achieve near DFT-level accuracy in the physical and chemical description of a system but at the much lower computational cost that is closer to classical force fields. These machine learned force fields enable studies of systems and mechanisms that have so far been out of the realm of computational studies, such as identifying the growth mechanism of amorphous carbon coatings\cite{gapcarbon} under deposition and the composition and activity of nanoclusters in aqueous solutions \cite{nncluster}. The implementations have now also made their way into established molecular dynamics software, where they can in some cases be used as a plug-and-play replacement for traditional force fields\cite{lammpsagni, lammpsnn, deepmdkit, lammpsquip}. 

Going one step up in the theoretical ladder, energies of very accurate, yet computationally very expensive coupled cluster theory calculations have been learned \cite{Bartok/etal:2017,Margraf/Reuter:2018,Smith/etal:2018,Wilkins/etal:2019}. Another example is the training of exchange-correlation functionals or direct potential-to-density mappings from density or wave-function based methods\cite{xc1, xc2}.  Such efforts are an exciting step in expanding existing theoretical knowledge to new time and size regimes in materials modelling.

In summary, machine learning and data-driven approaches are being applied in materials science. The wider the range of successful applications, the higher the acceptance of materials data infrastructures will be. New applications will, in turn, challenge established machine learning methods, which will have to be further developed to address these challenges. This creates a feedback loop with developments in computer and data science and ensures that our MUSE continues to learn.


\section{Stakeholder relations}
\label{sec:industry_uptake}

\begin{figure*}[t]
  \centering
    \includegraphics[width=0.95\textwidth, ]{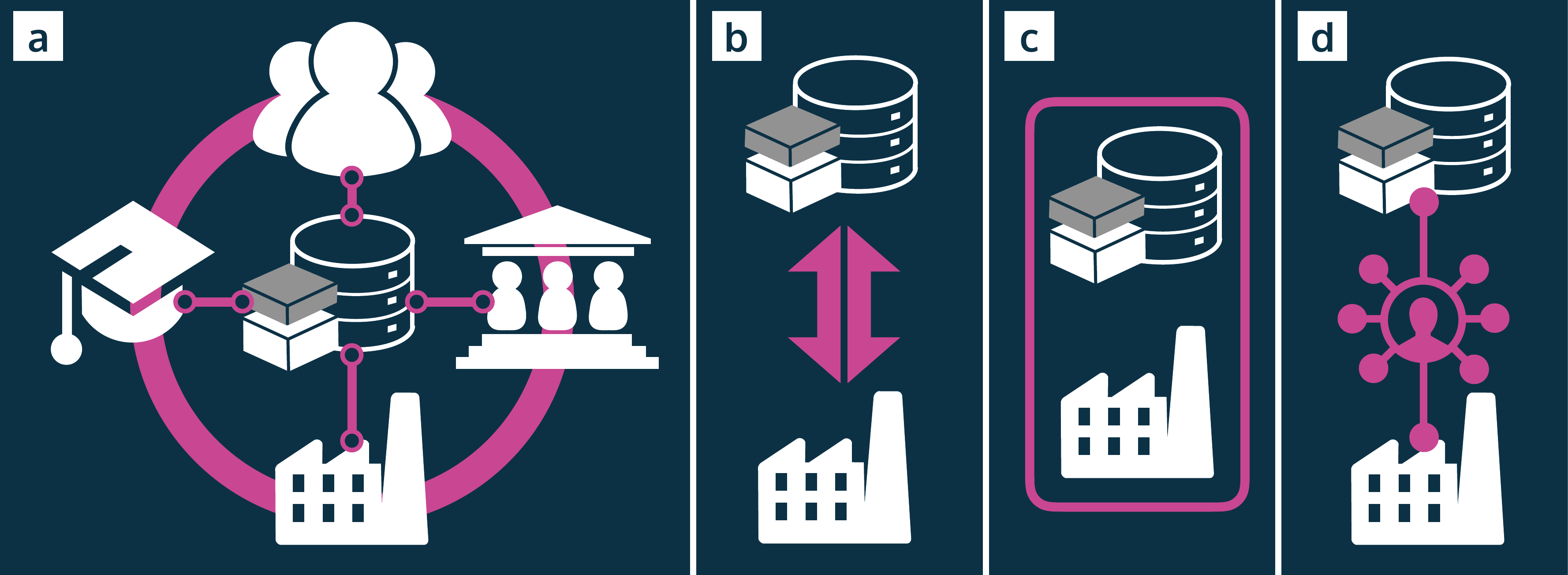}
  \caption{Schematic of an ecosystem in data-driven materials science with materials data platforms at the centre (panel a). In this ecosystem, different stakeholders from universities, the public, industry, and government facilitate the development of a technology. Panels b, c, and d highlight possible relationships between data platforms and industry and are discussed in the text.}
\label{fig:ecosystem}
\end{figure*}

In previous sections, we have addressed the \emph{acceptance} of materials data infrastructures from a technological viewpoint. We  now reflect on how materials data infrastructures are currently received by different stakeholders. Strong support from stakeholders is needed to guarantee the \emph{diffusion} of innovations to a wider pool of stakeholders and to ensure the \emph{longevity} of data infrastructures.

Materials scientists in academia are actively pushing the frontiers of materials informatics to advance and accelerate materials design and discovery. However emerging fields require the interaction of various actors and stakeholders from different communities (academia, government, industry, and the public, cf. Fig. \ref{fig:ecosystem} panel a) to generate understanding of the  field and negotiate community boundaries\cite{Granqvist}. A recent socio-economic study investigated the emerging field of data-driven materials science. It identified that the field is scattered and largely lacking a supportive ecosystem with non-academic stakeholders\cite{DataSciMat,Geurts/Granqvist/Rinke:2019}. 

The socio-economic study identified visionary scientists at academic institutions who have pursued their idiosyncratic research objectives using trending topics in public discussions and governmental funding, including \emph{Open Science}, \emph{Big Data}, and \emph{AI}. At the same time, government and funding agencies have provided strategic research openings focused on propelling the field of \emph{data-driven science} in general. However the focus on materials science applications or on finding and developing industry applications has been limited by the lack of targeted funding opportunities. With the exception of the US, which released substantial government funding to advance the field of data-driven materials science via the Materials Genome Initiative\cite{materialsgenome}, most government-sponsored funding schemes have been more general. In the European Union, two successful projects have been funded (MaX\cite{Max} and NOMAD CoE\cite{nomad}). However both centres were facilitated by the European Union’s Horizon 2020 high-performance computing grants rather than funding schemes focused on data-driven \emph{materials} science. Until now, no call tailored to the exploration and exploitation of data-driven materials science has been issued by the EU, although this may change in the new framework program. Nevertheless national differences exist. For example in Switzerland, the importance of data-driven materials science has been acknowledged by the support for the MARVEL National Center of Competence in Research (NCCR)\cite{mcurl}. In 2018 in Finland, a Future Makers funding call was opened specifically focused on the initiation of \emph{``high-level, ambitious strategic research openings that combine internationally top-level science and industrial impact ... to build long-term sustainable renewal and competitiveness of the Finnish technology industry based on ... data-based materials science''}\cite{Finland}. Despite the call, none of the applications from materials science were funded.  

The socio-economic study further concludes that industry has remained seemingly reluctant to invest in data-driven scientific applications in materials research and development. This reluctance could be driven by government hesitation to fund data-driven materials science, but it could also be the result of the material industries' desire to work with proprietary databases, the generally long timeframe for the development of new materials (10-20 years), and/or the limited number of manufacturing employees with informatics backgrounds\cite{Meredig}. Industries do, however, capitalize on the creation and appropriation of (new) knowledge\cite{Katila}, and new materials could advance such fields as health, energy, aerospace, automotive, semiconductor, and consumer goods\cite{Meredig}: materials informatics provides unprecedented opportunities for an industry to better use the existing vast \emph{``storehouses of information''}\cite{Garud} that firms possess to propel materials innovation at greater rates and lower costs. Despite the potential gains, uptake by industrial partners is challenging. Although materials companies generate enormous quantities of R\&D data, this data is often undocumented and intangible. Before proprietary databases could be created, the archival data of companies needs to be structured, connected, and updated. Ignoring the identified challenges - \emph{acceptance} (easy data upload and download, data curation, materials gap), \emph{standardization}, and \emph{longevity} - slows down industry adoption of data-driven materials science even further. In addition, firms face significant challenges to become or transform into data-driven organizations as they require different skills, knowledge, and resources\cite{Geurts}.   

To capitalize on their data, three business models can be applied or are considered for application\cite{DataSciMat,Geurts/Granqvist/Rinke:2019}. First, firms can consult and collaborate directly with established data platforms (see Fig. \ref{fig:ecosystem} panel b), for example those discussed in Figs.~\ref{fig:map_of_centres} and \ref{fig:centre_time_evolution}. Traditionally such collaborations take the form of strategic inter-firm alliances that influence companies' potential for knowledge creation\cite{Schilling}. Propelled by the drive for Open Science from policy players, Open Innovation is another potential form of collaboration in which firms open their internal innovation processes by purposefully allowing knowledge to freely circulate among all actors to accelerate internal innovation\cite{chesbrough}. As a result, such data platforms offer data and services that can be used by academia \emph{and} industries alike. 

In the second model, if a firm does not wish to collaborate, it can consider building a proprietary digital infrastructure by dedicating a large, one-off investment in hardware and by acquiring software and specialized skills in software configuration and system management\cite{Teece} (Fig. \ref{fig:ecosystem} panel c). This integration  requires attracting computer scientists or materials scientists with extensive coding capabilities. 

A final business model is developed around new intermediaries that position themselves between data platforms and industry (Fig. \ref{fig:ecosystem} panel d). Examples of such new intermediaries are materials informatics companies - often spin-off companies from academic efforts - that sell access to privately owned, highly curated materials property data that can be linked to data repositories within the firm and used in R\&D processes. Collaboration with start-ups and companies (e.g., Citrine Informatics, Exabyte.io, Materials Design, and Granta Design\cite{ citrineurl,exabyteurl,mdurl,grantaurl}) can help firms develop new skills, capabilities, and knowledge\cite{Schilling,gawer}.

Each of these data platform-industry relationships have implications for the larger ecosystem of data-driven materials science. That is, industry and academia often hold contradicting interests. From the academic perspective, commercial business opportunities stimulate private data ownership and proprietary databases (e.g., Fig. \ref{fig:ecosystem} panel c), which can be detrimental for scientific progress since valuable data stays locked in the private domain. Furthermore, industries outside materials science are quickly recruiting academic employees with new coding and machine learning expertise in the field; this raises concerns over a possible ``brain drain'' from universities. Nevertheless the establishment and institutionalization of data science and machine learning for materials science within educational institutions could result in an increase in research funding, students, and industry interest or collaboration\cite{DataSciMat}. 

At the same time, industries' need for these newly-skilled employees raises fear of ``job loss'' among established scientists within R\&D departments in those firms. However materials that are identified as promising through the materials informatics paradigm still require further evaluation, selection, experimentation, certification, and manufacturing. Building synergies among computational and experimental researchers therefore remains a key enabler toward reaping the benefits of data-driven materials science in firms\cite{DataSciMat}. 

In summary, the field of data-driven materials science is still in its infancy, with an emerging ecosystem, ongoing community boundary negotiations, limited governmental funding, and an as yet disinterested industry. To establish data-driven materials science as a new paradigm in materials research, joint ecosystem efforts between research, industry, and public and governmental organizations are necessary. 

\section{Take-home messages}

In this review article, we provide an overview of the current state of data-driven materials science. From a historical perspective, the field has matured greatly, but we identify key challenges - relevance, completeness, standardization, acceptance, and longevity - that still need to be resolved to create the Materials Ultimate Search Engine (MUSE).
Better standardization of materials data through a materials ontology would immensely help sharing, integrating, and employing AI-powered analysis of materials data. Creating feedback mechanisms between experimental and computational data for error estimation provides a way toward solving the veracity problem in materials data. The use of machine learning is transformational for research, but it requires conscious efforts in the curation and standardization of both data and machine learning models, and techniques to make the models more interpretable. The synergy between academic developments and industrial interest remains a major challenge, but it is key to creating a sustainable ecosystem for materials data and expertise.
Despite these challenges, there has been a dramatic rise in data-driven materials science using the full spectrum of this new paradigm. And doubtless we have only seen a glimpse of this data-driven revolution.

\section{Acknowledgements}
We thank Nina Granqvist, Kunal Ghosh, Ben Alldritt, Antti M. Rousi, Milica Todorovi{\'c}, Sven Bossuyt, Miguel Caro, David Gao, Matthias Scheffler, Bryce Meredig, and Heidi Henrickson for insightful discussions and a careful reading of our manuscript. Computing resources from the Aalto Science-IT project and CSC IT Center for Science, Finland, are gratefully acknowledged. This project has received funding from the European Union's Horizon 2020 research and innovation programme under grant agreement number 676580 with The Novel Materials Discovery (NOMAD) Laboratory, a European Center of Excellence, and from the Jenny and Antti Wihuri Foundation. This work was furthermore supported by the Academy of Finland through its Centres of Excellence Programme 2015-2017 under project number 284621, as well as projects 305632, 311012, and 314862. ASF was supported by the World Premier International Research Center Initiative (WPI), MEXT, Japan.


\bibliographystyle{advancedmaterials}
\bibliography{main}

\end{document}